\documentclass[manuscript,endfloat]{geophysics}
\usepackage{amsmath}
\usepackage{bm}
\usepackage{algorithm}
\usepackage{algpseudocode}

\begin{document}

\title{Numerical analysis of a deep learning formulation of elastic full waveform inversion with high order total variation regularization in different parameterization}
\renewcommand{\thefootnote}{\fnsymbol{footnote}} 

\address
{\footnotemark[1] University of Calgary, \\
	                         Department of Geoscience, \\
	                        Calgary, Alberta, Canada. \\
\footnotemark[2] Pennsylvania State University,\\
                            College of Earth and Mineral Science, \\
                            University Park, Pennsylvania, USA}
\author{Tianze Zhang\footnotemark[1]
	         Jian Sun\footnotemark[1]\footnotemark[2]
	         Kristopher A. Innanen\footnotemark[1]
	         Daniel Trad\footnotemark[1]}
\maketitle

\begin{abstract}
We have formulated elastic seismic full waveform inversion (FWI) within a deep learning environment.  Training this network with a single seismic data set is equivalent to carrying out elastic FWI. There are three main motivations for this effort.  The first is an interest in developing an inversion algorithm which is more or less equivalent to standard elastic FWI but which is ready-made for a range of cloud computational architectures.  The second is an interest in algorithms which can later (i.e., not in this paper) be coupled with a more involved training component, involving multiple datasets.  The third is a general interest in developing the idea of {\it theory-guiding} within machine learning solutions for large geophysical problems, wherein the number of degrees of freedom within a network, and the reliance on exhaustive training data, are both reduced by constraining the network with physical rules.  In our formulation, a recurrent neural network is set up with rules enforcing elastic wave propagation, with the wavefield projected onto a measurement surface acting as the synthetic data to be compared with observed seismic data.  Gradients for iterative updating of an elastic model, with a variety of parameterizations and misfit functionals, can be efficiently constructed within the network through the automatic differential method. With this method, the inversion based on complex misfits can be calculated.  We survey the impact of different complex misfits (based on the $l_{2}$, $l_{1}$) with high order total variation (TV) regulations on multiparameter elastic FWI recovery of models within velocity/density, modulus/density, and stiffness parameter/density parameterizations. We analyze parameter cross-talk. Inversion results on simple and complex models show that the RNN elastic FWI with high order TV regulation using $l_1$ norm can help mitigate cross-talk issues with gradient-based optimization methods.
\end{abstract}

\section{Introduction} 

It has recently been shown \citep[][]{sun2020theory} that seismic wave propagation can be simulated with a specialized recurrent neural network (RNN), and that the process of training such a network with a single seismic data set is equivalent to carrying out seismic full waveform inversion (FWI).  We are motivated to extend and expand on these results, because of (1) the apparent potential for wider training of such a network to combat common FWI issues such as modelling error; (2) the opportunities for efficient computation (e.g., cloud) offered by FWI realized on platforms like TensorFlow, and (3) an interest in the behavior of more complex RNN-FWI formulations than previously analyzed, e.g., multi-parameter elastic FWI.  In this paper we report on progress on the third of these.

The application of machine learning methods to seismic problems has been underway for decades; for example, \citet[][]{roth1994neural} presented a neural network-based inversion of time-domain seismic data for acoustic velocity depth-profiles.  However, the evolution of these network approaches into deep learning methods, and the results which have subsequently become possible, make many aspects of the discipline quite new, and explain the major recent surge in development and interest.   Now, novel seismic applications are being reported in fault detection, denoising, reservoir prediction and velocity inversion \citep[e.g.,][]{jin2019machine,zheng2019applications,peters2019neural,chen2019improving,li2019hybrid,smith2019correlating,shustak2018time}.  Specifically in seismic velocity inversion, \citet{sun2018low} applied a deep learning method to the problem of bandwidth extension; \cite{zhang2019velocitygan} designed an end-to-end framework, the velocity-GAN, which generates velocity images directly from the raw seismic waveform data; \citet{wu2018inversionnet} trained a network with an encoder-decoder structure to model the correspondence between seismic data and subsurface velocity structures; and \citet{yang2019deep} investigated a supervised, deep convolutional neural network for velocity-model building directly from raw seismograms. 

The above examples are purely data-driven, in the sense that they involve no assumed theoretical/physical relationships between the input layer (e.g., velocity model) and output layer (e.g., seismic data).  We believe that the advantages of the purely data-driven methods are that once the training for the network to perform inversion is finished, take the data-driven training for a network that can perform FWI as an example, the raw seismograms can be directly mapped to the velocity models. This would become a faster inversion method compared with the conventional inversion method that requires iterations. However,  seismic inversion is a sophisticated issue, so how we choose the sufficient amount of training data sets that can represent such complex wave physics features and their corresponding velocity models is a hard problem. 

Here we distinguish between such methods and those belonging to the {\it theory-guided} AI paradigm \citep[e.g.,][]{wagner2016theory,wu2018physics,karpatne2017theory}.  Theory-guided deep learning networks are, broadly, those which enforce physical rules on the relationships between variables and/or layers.  This may occur through the training of a standard network with simulated or synthetic data, which are created using physical rules, or by holding some weights in a network, chosen to force the network to mimic a physical process, fixed and un-trainable.  Theory-guiding was explicitly used in the former sense by \citet{downton18}, in which a network designed to predict well log properties from seismic amplitudes was trained not only with real data but with synthetics derived from the Zoeppritz equations.  \cite{bar2019learning} and \cite{raissi2018deep} built deep convolutional neural networks (or CNNs) to solve partial differential equations, i.e., explicitly using theoretical models within the design, which is an example of theory guiding in the latter sense.  \citet{sun2020theory}, in the work we extend in this paper, similarly set up a deep recurrent neural network to simulate the propagation of a seismic wave through a general acoustic medium.  The network is set up in such a way that the trainable weights correspond to the medium property unknowns (i.e., wave velocity model), and the non-trainable weights correspond to the mathematical rules (differencing, etc.) enforcing wave propagation.  The output layer was the wave field projected onto a measurement surface, i.e., a simulation of measured seismic data.  The training of the \citet{sun2020theory} network with a single data set was shown to be an instance of full waveform inversion.  

Conventional (i.e., not network-based) seismic full waveform inversion, or FWI, is a complex data fitting procedure aimed at extracting important information from seismic records. Key ingredients are an efficient forward modeling method to simulate synthetic data, and a local differential approach, in which the gradient and direction are calculated to update the model.  When updating several parameters simultaneously, which is a multiparameter full waveform inversion, error in one parameter tends to produce updates in others, a phenomenon referred to as inter-parameter trade-off or cross-talk \cite[e.g.,][]{kamei2013inversion,alkhalifah2014recipe,innanen2014seismic, pan2018elastic, keating2019parameter}.  The degree of trade-off or cross-talk between parameters can depend sensitively on the specific parameterization; even though, for instance, $\lambda$, $\mu$ and $\rho$ have the same information content as $V_P$, $V_S$, and $\rho$, updating in one can produce markedly different convergence properties and final models that doing so in the other.  The coupling between different physical parameters is controlled by the physical relationships amongst these parameters, and the relationships between the parameters and the wave which interacts with them. \cite{tarantola1986strategy}, by using the scattering or radiation pattern, systematically analyzed the effect of different model parameterizations on isotropic FWI. He suggested that the greater the difference in the scattering pattern between each parameter, the better the parameters can be resolved. \cite{kohn2012influence} showed that across all geophysical parameterizations, within isotropic-elastic FWI, density is always the most challenging to resolve \citep[e.g.,][]{tarantola1984inversion, plessix2006review,kamath20183d, lailly1983sequence, operto2013guided, oh2016elastic, pan2016estimation, keating2017crosstalk}.  

In RNN-based FWI, then, a natural step is the extension of the \citet{sun2020theory} result, which involves a scalar-acoustic formulation of FWI, to a multi-parameter elastic version.  Cells within a deep recurrent neural network are designed such that the propagation of information through the network is equivalent to the propagation of an elastic wavefield through a 2D heterogeneous and isotropic-elastic medium; the network is equipped to explore a range of parameterizations and misfit functionals for training based on seismic data.  As with the acoustic network, the output layer is a projection of the wave field onto some defined measurement surface, simulating measured data.

In addition to providing a framework for inversion methods which mixes the features of FWI with the training capacity of a deep learning network, this approach also allows for efficient calculation of the derivatives of the residual through automatic differential (AD) methods \citep[e.g.,][]{li2019time,sambridge2007automatic}, an engine for which is provided by the open-source machine learning library Pytorch \citep{paszke2017automatic}.  Our recurrent neural network is designed using this library.  The forward simulation of wave propagation is represented by a Dynamic Computational Graph (DCG), which records how each internal parameter is calculated from all previous ones. In the inversion, the partial derivatives of the residual with respect to any parameter (within a trainable class) is computed by (1) backpropagating within the computational graph to that parameter, and (2) calculating the partial derivative along the path using the chain rule.  

This paper is organized as follows. First, we introduce the basic structure of the recurrent neural network and how the gradients can be calculated using the backpropagation method. Second, we demonstrate how the elastic FWI RNN cell is constructed in this paper and how the waveflieds propagate in the RNN cells. Third, we explain the $l_2$ and $l_1$ misfits with high order TV regulation and the mathematical expression for the gradients based on these misfits. Fourth, we use simple layers models and complex over-thrust models to perform inversions with various parameterizations using different misfits. Finally, we discuss the conclusions of this work. 

\section{Recurrent Neural Network}

A recurrent neural network (RNN) is a machine learning network suitable for dealing with data which have some sequential meaning. Within such a network, the information generated in one cell layer can be stored and used by the next layer.  This design has natural applicability in the processing and interpretation of the time evolution of physical processes  \citep[][]{sun2020theory}, and time-series data in general; examples include language modeling \citep[][]{mikolov2012statistical}, speech recognition \cite[][]{graves2013speech}, and machine translation \citep[][]{kalchbrenner2013recurrent,zaremba2014recurrent}.   

\begin{figure}[h!]
\centering
\includegraphics[width=1\textwidth]{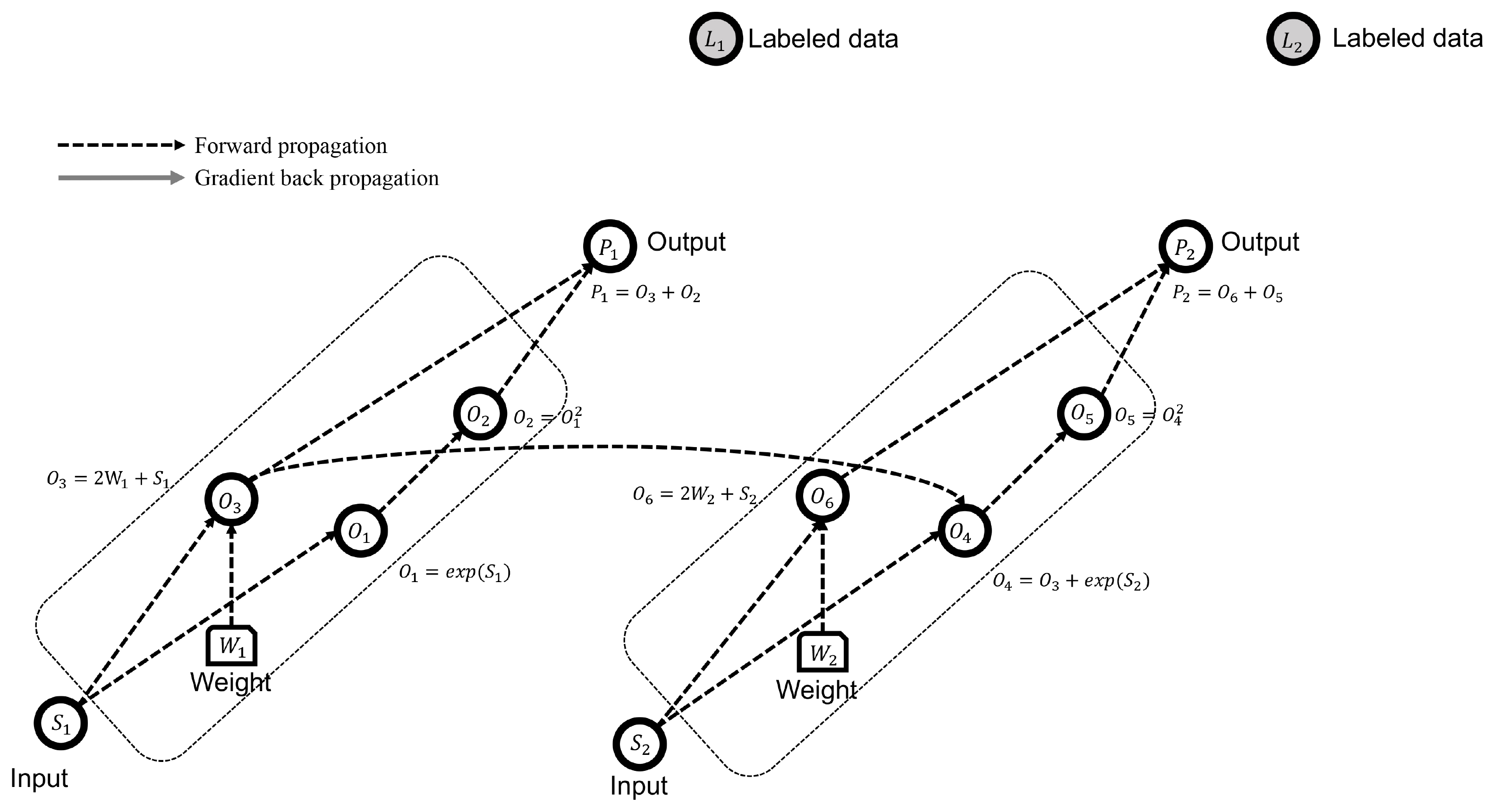}
\caption{Forward propagation through an example RNN.  Cells are identical in form, with the output of one being the input of the next.  $\textbf{O}=[O_{1}, O_{2}, O_{3},{\cdots}]$ are the internal variables in each cell; $\textbf{S}=[S_{1}, S_{2}]$  are the inputs; $\textbf{P}=[P_{1}, P_{2}]$ are the outputs; $\textbf{L}=[L_{1}, L_{2}]$ are the labeled data; $\textbf{W}=[W_{1}, W_{2}]$ are the trainable weights. Black dashed line indicates forward propagation.}\label{fig:RNN_forward.pdf}
\end{figure}
\begin{figure}[h!]
\centering
\includegraphics[width=1\textwidth]{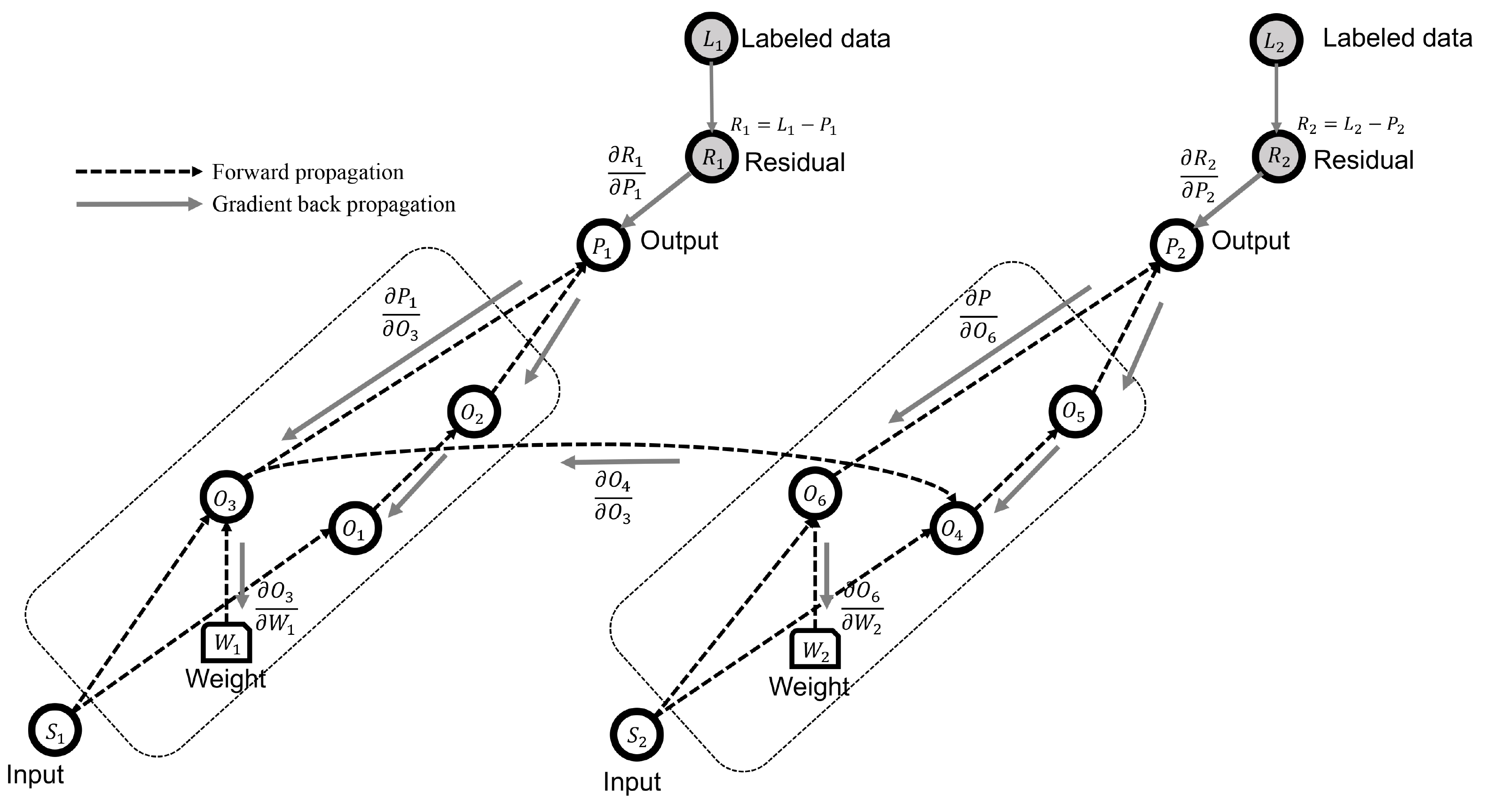}
\caption{Residual backward propagation through an example RNN. $\textbf{R}=[R_{1}, R_{2}]$   are the residuals at each RNN cell, which are calculated using absolute error($l_1$ norm). Gray solid line shows how gradients are calculated using back propagation from the residual to the trainable weights in RNN cell along the computational graph. }\label{fig:RNN_backward.pdf}
\end{figure}

Figure \ref{fig:RNN_forward.pdf} illustrates the forward propagation of information through an example RNN, in which each RNN cell represents an instant in time: the $\textbf{O}=[O_{1}, O_{2}, O_{3}, {\cdots}]$ are the internal variables in each RNN cell; $\textbf{S}=[S_{1}, S_{2}]$ are the input at each time step; $\textbf{P}=[P_{1}, P_{2}]$ are the output; $\textbf{L}=[L_{1}, L_{2}]$ are the labeled data; and $\textbf{W}=[W_{1}, W_{2}]$ are the trainable parameters at each time step.  Mathematical operations relating the internal variables within a cell, and those relating internal variables across adjacent cells, are represented as arrows.  

To train this network is to select trainable weights such that labeled data ${\textbf{L}}$ and RNN output ${\textbf{P}}$ are as close as possible. Specifically, in training we determine the parameters ${\textbf{W}}$, through a gradient-based optimization involving the partial derivatives of the residual with respect to each $W_i$. These derivatives are determined through backpropagation, which is a repeated application the chain rule, organized to resemble flow in the reverse direction along with the arrows in the network.  For the example RNN in Figure \ref{fig:RNN_forward.pdf}, this takes the form illustrated in Figure \ref{fig:RNN_backward.pdf}, in which the sequence $\textbf{R}=[R_{1}, R_{2}]$ represents the residuals at each time step.  Within this example, to calculate the partial derivative of $R_{1}$ with respect to $W_{1}$, we back-propagate from node $R_{1}$ to $W_{1}$: 
\begin{equation}
\frac{{\partial}R_{1}}{{\partial}W_{1}}=\frac{{\partial}R_{1}}{{\partial}P_{1}}\frac{{\partial}P_{1}}{{\partial}W_{1}}=
\frac{{\partial}R_{1}}{{\partial}P_{1}}\frac{{\partial}P_{1}}{{\partial}O_{3}}\frac{{\partial}O_{3}}{{\partial}W_{1}}=-2
\end{equation} To calculate the partial derivative of $R_{2}$ with respect to $W_{1}$, we backpropagate from $R_{2}$ to $W_{1}$:
\begin{equation}
\frac{{\partial}R_{2}}{{\partial}W_{1}}=\frac{{\partial}R_{2}}{{\partial}P_{2}}\frac{{\partial}P_{2}}{{\partial}W_{1}}=
\frac{{\partial}R_{2}}{{\partial}P_{2}}\frac{{\partial}P_{2}}{{\partial}O_{5}}\frac{{\partial}O_{5}}{{\partial}O_{4}}
\frac{{\partial}O_{4}}{{\partial}O_{3}}\frac{{\partial}O_{3}}{{\partial}W_{1}}=-4O_{4}
\end{equation} If the RNN was set up to propagate through two-time steps, the gradient for $W_{1}$ is $-4O_{4}+2$.  Real RNNs are more complex and involve propagation through larger numbers of time steps, but all are optimized through a process similar to this.  These derivatives are the basis for gradients in the optimization misfit function; using them the ${\textbf{W}}$ are updated and iterations continue. 
\section{A Recurrent Neural Network formulation of EFWI}

Wave propagation can be simulated using suitably-designed RNNs \citep[][]{sun2020theory,richardson2018seismic,hughes2019wave}. Here we take the acoustic wave propagation approach of \citet{sun2020theory} as a starting point, and formulate an RNN which simulates the propagation of an elastic wave through isotropic elastic medium.  The underlying equations are the 2D velocity-stress form of the elastodynamic equations \citep[][]{virieux1986p,liu2009implicit}:
\begin{equation}	
\begin{aligned}
&{\frac{\partial {{v}}_{x}}{\partial t}  =
\frac{1}{\rho}\left(\frac{\partial {\sigma}_{x x}}{\partial x}+\frac{\partial {\sigma}_{x z}}{\partial z}\right)} \\ 
&{\frac{\partial {{v}}_{z}}{\partial t}  =
\frac{1}{\rho}\left(\frac{\partial {\sigma}_{x z}}{\partial x}+\frac{\partial {\sigma}_{z z}}{\partial z}\right)} \\ 
&{\frac{\partial {\sigma}_{x x}}{\partial t} =
({\lambda}+2{\mu}) \frac{\partial {{v}}_{x}}{\partial x}+{\lambda}\frac{\partial {{v}}_{z}}{\partial z}} \\ 
&{\frac{\partial {\sigma}_{z z}}{\partial t}  =
({\lambda}+2{\mu}) \frac{\partial {{v}}_{z}}{\partial z}+{\lambda}\frac{\partial {{v}}_{x}}{\partial x}} \\ 
&{\frac{\partial {\sigma}_{x z}}{\partial t}  =
{\mu}\left(\frac{\partial {{v}}_{x}}{\partial z}+\frac{\partial {{v}}_{z}}{\partial x}\right)} \\
\end{aligned}. ,
\end{equation} where ${{v}}_{x}$ and ${{v}}_{z}$ are the $x$ and $z$ components of the particle velocity, ${{\sigma}_{x x}}$, ${{\sigma}_{z z}}$ and ${{\sigma}_{x z}}$ are three 2D components of the stress tensor.   Discretized spatial distributions of the Lam\'{e} parameters $\lambda$ and $\mu$, and the density ${\rho}$, form the elastic model. 

\begin{figure}[h!]
\centering
\includegraphics[width=1\textwidth]{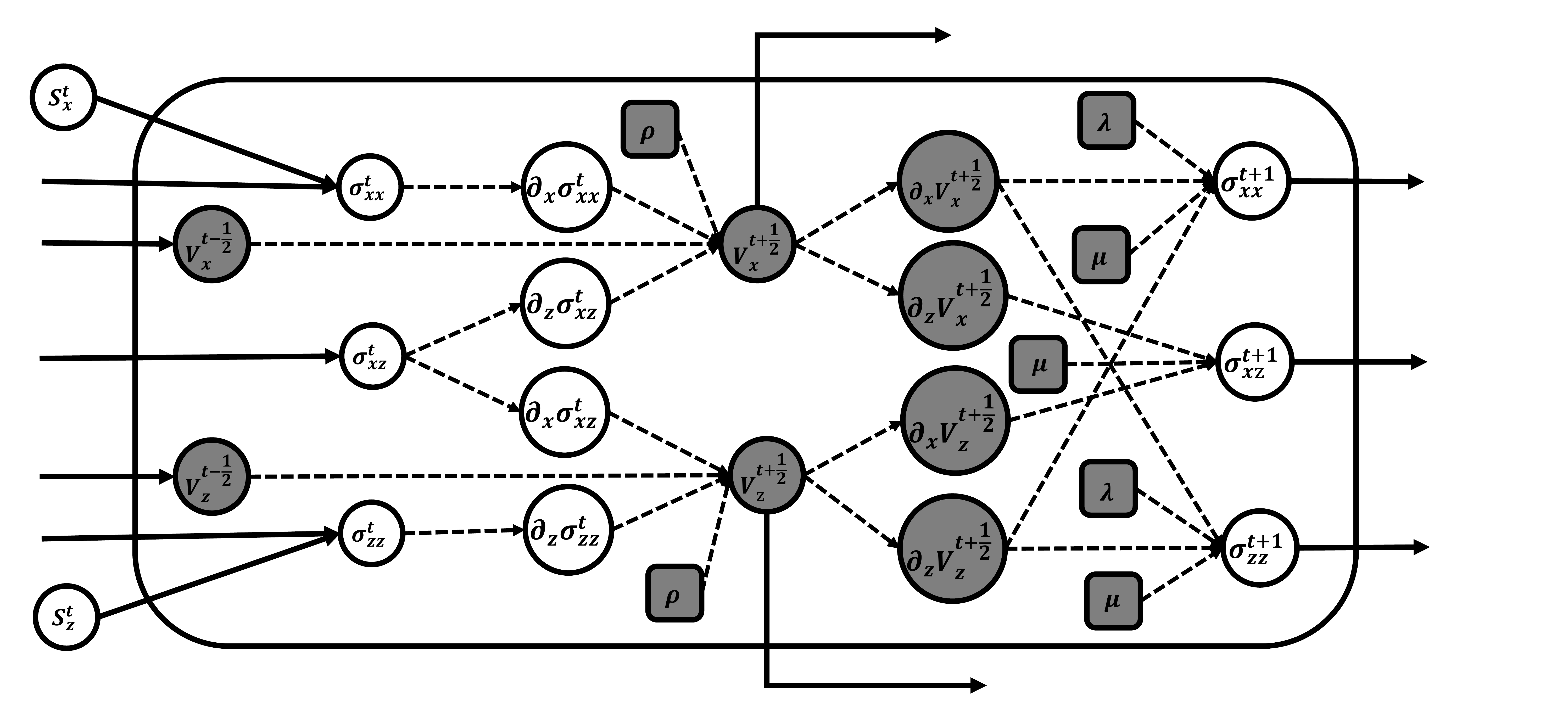}
\caption{The structure of each RNN. In this figure, 
	$\partial_{x}{\sigma}_{xx}^{t}$, 
	$\partial_{z}{\sigma}_{zz}^{t}$,
	$\partial_{x}{\sigma}_{xz}^{t}$, 
	$\partial_{z}{\sigma}_{xz}^{t}$,
	$\partial_{x}{v}_{x}^{t+\frac{1}{2}}$, 
	$\partial_{z}{v}_{x}^{t+\frac{1}{2}}$,
	$\partial_{x}{v}_{z}^{t+\frac{1}{2}}$, 
	$\partial_{z}{v}_{z}^{t+\frac{1}{2}}$ are the internal variables. 
		${v}^{t-\frac{1}{2}}_{x}$, 
	${v}^{t-\frac{1}{2}}_{z}$, 
	${\sigma}_{xx}^{t}$, 
	${\sigma}_{zz}^{t}$, 
	${\sigma}_{xz}^{t}$, 
	 is communicated between the RNN cells. 
	 ${\lambda}$, ${\mu}$,$\rho$ are trainable parameters. } \label{fig:Elastic_RNN_cell.pdf}
\end{figure}
\begin{algorithm}[h]
	\caption{Sequence of calculations in the RNN cell with PML boundary}
	\label{alg:Framwork}
	\begin{algorithmic}[1]
		\Require
		Source: $s_{x}$, $s_{z}$; 
		velocity and stress fields at the previous time step, parameters :$\lambda$, $\mu$, $\rho$; time step: $dt$. PML damping coefficients $d_x$. $d_z$.
		\Ensure
		Update velocity field at $t+\frac{1}{2}$ and stress fields at $t+1$
		\State ${\sigma_{x x}^{t}}  \leftarrow {\sigma_{x x}^{t}}+{s}_{x} $
		\State ${\sigma_{z z}^{t}}  \leftarrow {\sigma_{z z}^{t}}+{s}_{z} $
		\State $\partial_{x}{\sigma_{x x}^{t}}  \leftarrow ({\sigma_{x x}^{t}} * {\mathbf k}_{x_1}) / \rho $
		\State $\partial_{z}{\sigma_{x z}^{t}}  \leftarrow ({\sigma_{x z}^{t}} * {\mathbf k}_{z_2}) / \rho $
		\State $\partial_{x}{\sigma_{x z}^{t}}  \leftarrow ({\sigma_{x z}^{t}} * {\mathbf k}_{x_2}) / \rho $
		\State $\partial_{z}{\sigma_{z z}^{t}}  \leftarrow ({\sigma_{z z}^{t}} * {\mathbf k}_{z_1}) / \rho $
		\State ${{v}_{x}^{t+\frac{1}{2}}}_{x} 
		\leftarrow (1 - dt d_{x} ){{v}_{x}^{t-\frac{1}{2}}}_{x} +
		dt(\partial_{x} {\sigma_{x x}^{t}})$
		\State ${{v}_{x}^{t+\frac{1}{2}}}_{z} 
		\leftarrow (1 - dt d_{z} ){{v}_{x}^{t-\frac{1}{2}}}_{z} +
		dt(\partial_{z} {\sigma_{x z}^{t}})$
		\State ${v}_{x}^{t+\frac{1}{2}} 
		\leftarrow {{v}_{x}^{t+\frac{1}{2}} }_{x} + {{v}_{x}^{t+\frac{1}{2}} }_{z}$  
		\State ${{v}_{z}^{t+\frac{1}{2}}}_{x} 
		\leftarrow (1 - dt d_{x} ){{v}_{z}^{t-\frac{1}{2}}}_{x} +
		dt(\partial_{x} {\sigma_{x z}^{t}})$
		\State ${{v}_{z}^{t+\frac{1}{2}}}_{z} 
		\leftarrow (1 - dt d_{z} ){{v}_{z}^{t-\frac{1}{2}}}_{z} +
		dt(\partial_{z} {\sigma_{z z}^{t}})$
		\State ${v}_{z}^{t+\frac{1}{2}} 
		\leftarrow {{v}_{x}^{t+\frac{1}{2}} }_{x} + {{v}_{x}^{t+\frac{1}{2}} }_{z}$
		\State $\partial_{x}{v}_{x}^{t+\frac{1}{2}} \leftarrow {v}_{x}^{t+\frac{1}{2}} * {\mathbf k}_{x_2}$
		\State $\partial_{z}{v}_{x}^{t+\frac{1}{2}} \leftarrow {v}_{x}^{t+\frac{1}{2}} * {\mathbf k}_{z_1}$
		\State $\partial_{x}{v}_{z}^{t+\frac{1}{2}} \leftarrow {v}_{z}^{t+\frac{1}{2}} * {\mathbf k}_{x_1}$
		\State $\partial_{z}{v}_{z}^{t+\frac{1}{2}} \leftarrow {v}_{z}^{t+\frac{1}{2}} * {\mathbf k}_{z_2}$
		\State ${{\sigma_{x x}}^{t+1}}_{x}  \leftarrow 
		(1 - dt dx){{\sigma_{x x}}^{t}}_{x} + 
		dt({\lambda+2\mu})\partial_{x}{v}_{x}^{t+\frac{1}{2}}$
		\State ${{\sigma_{z z}}^{t+1}}_{x}  \leftarrow 
		(1 - dt dz){{\sigma_{z z}}^{t}}_{x} +  
		dt(\lambda)\partial_{z}{v}_{z}^{t+\frac{1}{2}}$
		\State ${{\sigma_{z z}}^{t+1}}  \leftarrow 
		{{\sigma_{x x}}^{t+1}}_{x} +{{\sigma_{x x}}^{t+1}}_{z} $
		\State ${{\sigma_{z z}}^{t+1}}_{x}  \leftarrow 
		(1 - dt dx){{\sigma_{x x}}^{t}}_{x} + 
		dt({\lambda})\partial_{x}{v}_{x}^{t+\frac{1}{2}}$
		\State ${{\sigma_{z z}}^{t+1}}_{z}  \leftarrow 
		(1 - dt dz){{\sigma_{z z}}^{t}}_{z} +  
		dt(\lambda+2\mu)\partial_{z}{v}_{z}^{t+\frac{1}{2}}$
		\State ${{\sigma_{zz}}^{t+1}}  \leftarrow 
		{{\sigma_{z z}}^{t+1}}_{x} +{{\sigma_{z z}}^{t+1}}_{z} $
		\State ${{\sigma_{x z}}^{t+1}}_{x}  \leftarrow 
		(1 - dt dx){{\sigma_{x z}}^{t}}_{x} + 
		dt({\mu})\partial_{z}{v}_{x}^{t+\frac{1}{2}}$
		\State ${{\sigma_{x z}}^{t+1}}_{z}  \leftarrow 
		(1 - dt dz){{\sigma_{x z}}^{t}}_{z} +  
		dt(\mu)\partial_{x}{v}_{z}^{t+\frac{1}{2}}$
		\State ${{\sigma_{xz}}^{t+1}}  \leftarrow 
		{{\sigma_{xz}}^{t+1}}_{x} +{{\sigma_{xz}}^{t+1}}_{z} $								
	\end{algorithmic}
\end{algorithm}
 
In Figure \ref{fig:Elastic_RNN_cell.pdf} the structure of an RNN cell which produces a staggered-grid finite difference solution for the velocity and stress fields is illustrated.  At each time step the discrete sources $s_{x}$ and $s_{z}$ act as inputs; the velocity and stress information, ${v}^{t-\frac{1}{2}}_{x}$, ${v}^{t-\frac{1}{2}}_{z}$, ${\sigma}_{xx}^{t}$, ${\sigma}_{zz}^{t}$, and ${\sigma}_{xz}^{t}$, is communicated between the RNN cells ; the partial derivative fields,  
$\partial_{x}{\sigma}_{xx}^{t}$, $\partial_{z}{\sigma}_{zz}^{t}$,
$\partial_{x}{\sigma}_{xz}^{t}$, $\partial_{z}{\sigma}_{xz}^{t}$,
$\partial_{x}{v}_{x}^{t+\frac{1}{2}}$, $\partial_{z}{v}_{x}^{t+\frac{1}{2}}$,
$\partial_{x}{v}_{z}^{t+\frac{1}{2}}$, $\partial_{z}{v}_{z}^{t+\frac{1}{2}}$ are the internal variables in each RNN cell, which correspond to O in Figure \ref{fig:RNN_backward.pdf}; and, $\lambda$, $\mu$ and $\rho$ are included as trainable weights, which correspond to W in Figure \ref{fig:RNN_backward.pdf}. In Algorithm 1 pseudocode detailing these calculations within the RNN cell is provided.  The $*$ symbol represents the machine learning image convolution operator. This image convolution is the process of adding each element of the image to its local neighbors, weighted by the image convolution kernel. We find that this image convolution operator is also capable of calculating space partial derivatives if the convolution kernel is designed according to the finite difference coefficients. Details about the image convolution operation can be found in \cite{podlozhnyuk2007image}.  $dx$, $dz$ are the grid intervals, and the image convolution kernels are: ${\mathbf k}_{x_1}={\mathbf a}/dx$, ${\mathbf k}_{x_2}={\mathbf b}/dx$, ${\mathbf k}_{z_1}={\mathbf a}^T/dz$, and ${\mathbf k}_{z_2}={\mathbf b}^T/dz$, where ${\mathbf a} = [0,1/24,-9/8,9/8,-1/24]$ and ${\mathbf b} = [1/24,-9/8,9/8,-1/24,0]$.  ${\mathbf a}$ and ${\mathbf b}$ are 1$\times$5 dimension arrays.  ${\mathbf k}_{x_1}$ and ${\mathbf k}_{x_2}$ are kernels, for the image convolution process, responsible for calculating the staggered grid space partial derivative in x direction. ${\mathbf k}_{z_1}$ and ${\mathbf k}_{z_2}$ are kernels, for the image convolution process, responsible for calculating the staggered grid space partial derivative in z direction, and that is also why the arrays, ${\mathbf a}$ and ${\mathbf b}$,  are transposed in ${\mathbf k}_{z_1}$ and ${\mathbf k}_{z_2}$. Space partial derivative calculated in this way is, mathematically, the same with conventional staggered grid method (e.g.\cite{virieux1986p}). In this study, we achieve this process in an image convolution way.  In algorithm 1, in order to implement the PML boundary, all the stress fields and the velocity fields need to be split into their x and z components.   
In algorithm 1, $d_{x}$ and $d_{z}$ are the PML damping coefficients in x direction and z direction. $d_{x}$ can be expressed as:
\begin{equation}
d_{x}(i) = d_{0x}(\frac{i}{n_{pmlx}})^{p}
\end{equation}
,where * represents either x or z direction. i is PML layer number starting from the effective calculation boundary. $n_{pmlx}$ is the PML layer number in $x$ direction. p is an integer and the value is from 1-4. $d_{0x} $ can be expressed as: 
\begin{equation}
d_{0x} = log(\frac{1}{R})\frac{r{V_s}}{n_{pmlx}{\delta}_{x}}
\end{equation} 
,where $R$ is a theoretically reflection coefficient, $r$ is a value ranging from 3-4. ${\delta}_{x}$ is the grid length in $x$ direction.   $d_{z}$ can also be calculated in the same way. 

\begin{figure}[h!]
\centering
\includegraphics[width=1\textwidth]{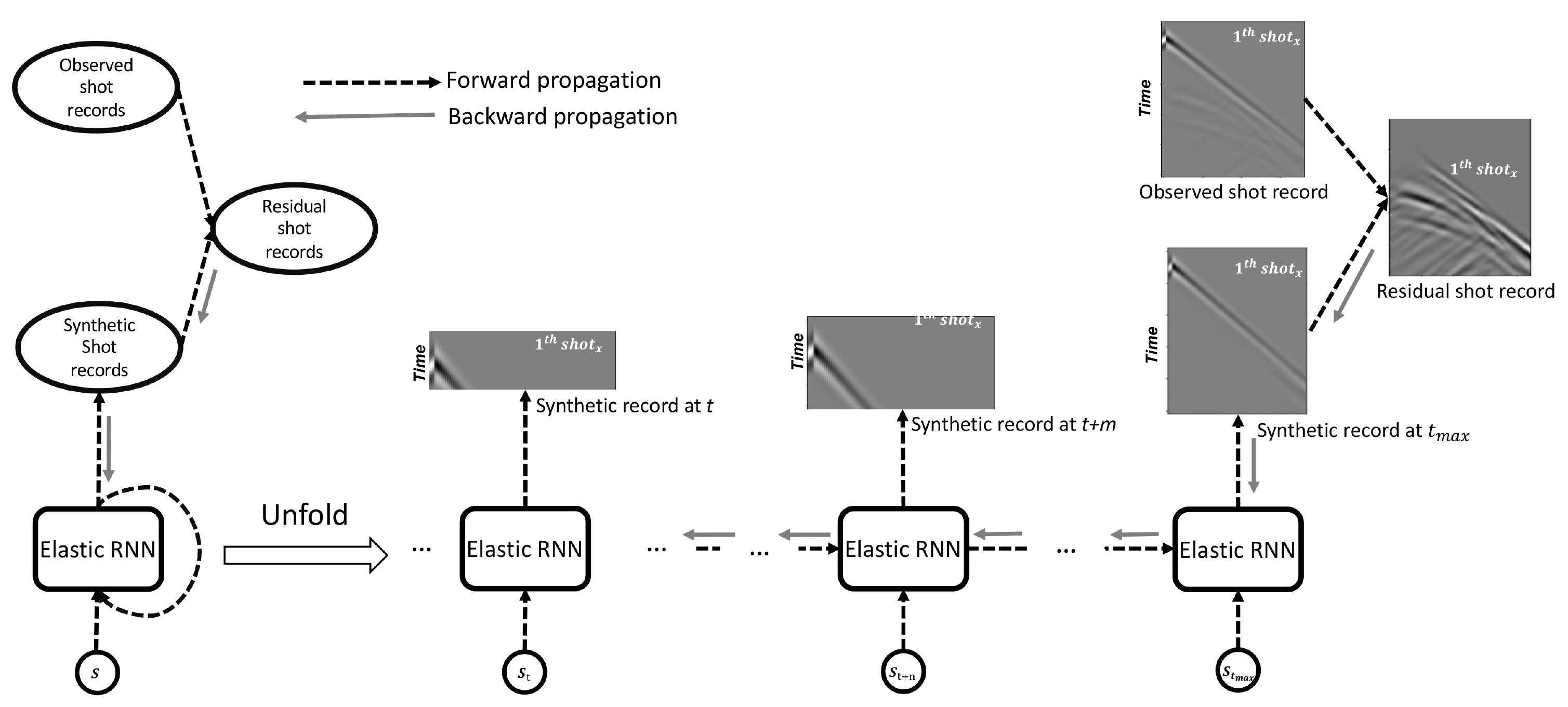}
\caption{Velocity field forward and residual backward propagation under formatting of the RNN. The shot records formed at each time step correspond to the output of RNN cell P in Figure \ref{fig:RNN_backward.pdf}. Observed shot records  correspond to labeled data (L in Figure \ref{fig:RNN_backward.pdf}). Residual shot record correspond to residual information (R in Figure \ref{fig:RNN_backward.pdf}). Back propagation starts from the residual shot record. }\label{fig:Elastic_RNN_propagate.pdf}
\end{figure}

The activity of the RNN network is illustrated in ``unfolded'' form in Figure \ref{fig:Elastic_RNN_propagate.pdf}.  Above the unfolded network, horizontal velocity fields associated with a point source at the top left of a model are plotted at three times during propagation of the wave information through the network, the third being at $t_{max}$, the maximum receiving time.  The wave field values are stored at positions selected to match multicomponent receivers; these form shot records as time evolves, which becomes the output data at each time. These shot records correspond to variables P in Figures \ref{fig:RNN_forward.pdf} and \ref{fig:RNN_backward.pdf}, and the observed data. For FWI problem, the observed data is obtained from the true model.  For real seismic data inversion, the observed data is obtained from field survey. In this RNN based elastic FWI the observed data is considered as the labeled data. The residuals are calculated at the last computational time as Figure \ref{fig:Elastic_RNN_propagate.pdf} illustrated and this along with a selected norm defines the misfit function used to train the network. 

Algorithm 2 describes the training of the network, i.e., the process of elastic RNN FWI.  Partial derivatives of the residual with respect to the trainable parameters (in this case $\lambda$, $\mu$ and $\rho$) are calculated through backpropagation using the automatic differential method, as set out in the previous section.  After we have the gradients then we can use an optimization method and step length to update the trainable parameters and reduce the misfit and start another iteration.  In step 4, RNN() is the network discussed above, whose output is the synthetic data; costFunc(), in step 5, is the misfit or loss function chosen to measure the difference between the synthetic data and observed data; loss.backward() begins the backpropagation within the computational graph and produces the gradients for each parameter, with which the current parameter model can be updated and another iteration started. 

\begin{algorithm}[h]
\caption{Loop for elastic RNN FWI}
\begin{algorithmic}[1]
\State Set trainable parameters: $\lambda$, $\mu$, ${\rho}$ in this test.
\State Set optimizers for parameters: $Optimizer_{1}$, $Optimizer_{2}$ and $Optimizer_{3}$ for ${\lambda}$, $\mu$ and ${\rho}$ respectively.
\For{ iter $\in [1,maxiter]$ or not converge}
\State $D_{syn}$ = RNN($\lambda$, $\mu$, ${\rho}$): generate synthetic data  
\State loss = costFunc($D_{syn}$, $D_{obs}$): calculate misfits  
\State loss.backward(): Backpropagation and give gradients for the parameters
\State optimizers.step(): update parameters 
\EndFor
\label{code:recentEnd}
\end{algorithmic}
\end{algorithm}  

The gradient calculated within the training process above essentially reproduces the adjoint-state calculations within FWI, as discussed by \citet{sun2020theory}.  Formulated as a problem of RNN training, the gradient calculation occurs rapidly and in a manner suitable for cloud computational architectures; also, it allows the researcher to efficiently alter misfit function choices and parameterization in order to do high-level optimization.  However, it involves the storage of the whole wavefield, and thus should be expected to have significant memory requirements. 

\section{Misfits with high order TV regulation}

Here we fist introduce the elastic RNN misfits based on $l_2$ norm with high order TV regularization:
\begin{equation}
\begin{aligned}
 {\mathbf {\Phi}}_{l2}^{TV}
 ({\mathbf {m_{\lambda},{m_{\mu},m_{\rho}}}},
 {\alpha_{1}^{\lambda}}, 
   {\alpha_{1}^{\mu}}, 
  {\alpha_{1}^{\rho}}, 
  {\alpha_{2}^{\lambda}}, 
  {\alpha_{2}^{\mu}}, 
  {\alpha_{2}^{\rho}})  
 = 
 \frac{1}{2} 
 \lVert \mathbf {D_{syn}} \mathbf {(m_{\lambda},m_{\mu},m_{\rho})}-\mathbf {D_{obs}} \rVert ^2_2 +\\
  {\alpha_{1}^{\lambda}}  \mathbf {{\Theta}_{TV}} \mathbf {(m_{\lambda})}  +
  {\alpha_{1}^{\mu}}  \mathbf {{\Theta}_{TV}} \mathbf {(m_{\mu})}  +
   {\alpha_{1}^{\rho}}  \mathbf {{\Theta}_{{TV}}} \mathbf {(m_{\rho})}  + \\
  {\alpha_{2}^{\lambda}}  \mathbf {\Upsilon_{TV}} \mathbf {(m_{\lambda})}  +
  {\alpha_{2}^{\mu}}  \mathbf {\Upsilon_{TV}}  \mathbf {(m_{\mu})}  +
  {\alpha_{2}^{\rho}}  \mathbf {\Upsilon_{TV}} \mathbf {(m_{\rho})} 
 \end{aligned}
\end{equation}
,where 
$ {\alpha_{1}^{\lambda}}$,
$ {\alpha_{1}^{\mu}}$,
$ {\alpha_{1}^{\rho}}$,
$ {\alpha_{1}^{\lambda}}$,
$ {\alpha_{1}^{\mu}}$,
$ {\alpha_{1}^{\rho}}$, are vector of Lagrange multipliers, and $\mathbf {{\Theta}_{TV}}$, $\mathbf {{\Upsilon}_{TV}} $ represents first and second order TV regularization functions respectively.   $\mathbf {D_{syn}} \mathbf {(m_{\lambda},m_{\mu},m_{\rho})}$ represents the synthetic data, which is the function of the model parameters, and in this equation they are  $V_P$, $V_S$ and $\rho$. $\mathbf {{\Theta}_{TV}}$ and $\mathbf {{\Upsilon}_{TV}}$ represent functions for calculating the first and second order TV regulations for the models. 

The first order TV regulation term can be expressed as: 
\begin{equation}
\begin{aligned}
TV_{1}( \mathbf {(m)} ) = 
& \sum_{i=1}^{n-1}\sum_{j=1}^{m-1}
| M_{i+1,j} - M_{i,j} | +\sum_{i=1}^{n-1}\sum_{j=1}^{m-1}
| M_{i,j+1} - M_{i,j} | =  \\
&  \begin{pmatrix} 
{\nabla}_{x},
{\nabla}_{z} 
\end{pmatrix} 
\begin{pmatrix} 
\mathbf {m}, \\
\mathbf {m}
\end{pmatrix} 
=
\begin{pmatrix} 
\mathbf {\mathcal L_{x}} ,
\mathbf {\mathcal L_{z}} ,
\end{pmatrix} 
\begin{pmatrix} 
\mathbf {m}, \\
\mathbf {m}
\end{pmatrix} 
=
\mathbf {{\Theta}_{TV}} \mathbf {(m)}
\end{aligned} 
\end{equation}

The second order TV regulation term can be expressed as:  
\begin{equation}
\begin{aligned}
TV_{2}( \mathbf {(m)} ) = 
& \sum_{i=1}^{n-1}\sum_{j=1}^{m-1}
| M_{i+1,j} - 2M_{i,j} +M_{i-1,j}| +
\sum_{i=1}^{n-1}\sum_{j=1}^{m-1}
| M_{i,j+1} - 2M_{i,j} + M_{i,j-1} |  \\
= 
&  \begin{pmatrix} 
{\nabla}_{xx},
{\nabla}_{zz} 
\end{pmatrix} 
\begin{pmatrix} 
\mathbf {m}, \\
\mathbf {m}
\end{pmatrix} 
=
\begin{pmatrix} 
 \mathbf {\mathcal K_{xx}},
 \mathbf {\mathcal K_{zz}} ,
\end{pmatrix} 
\begin{pmatrix} 
\mathbf {m}, \\
\mathbf {m}
\end{pmatrix} 
=
\mathbf {{\Upsilon}_{TV}} \mathbf {(m)}
\end{aligned} 
\end{equation}

The derivative of ${\mathbf {\Phi}}_{l2}^{TV}$ for each parameter ,which is the gradient for $V_P$, $V_S$ and $\rho$ based on the $l_2^{TV}$ norm can be expressed as: 
\begin{equation}
\begin{aligned}
\begin{pmatrix} 
\frac{{\partial}  {\mathbf {\Phi}}_{l2}^{TV} }{{\partial} {\mathbf{m_{\lambda}}}}\\
\frac{{\partial}  {\mathbf {\Phi}}_{l2}^{TV} }{{\partial} {\mathbf{m_{\mu}}}}\\
\frac{{\partial}  {\mathbf {\Phi}}_{l2}^{TV} }{{\partial} {\mathbf{m_{\rho}}}}\\
\end{pmatrix} = 
\begin{pmatrix}
\mathbf {G_{{l2}_{\lambda}}} \\
\mathbf {G_{{l2}_{\mu}}} \\
\mathbf {G_{{l2}_{\rho}}} \\
\end{pmatrix}+
\begin{pmatrix}
\mathbf {R_{\lambda}} \\
\mathbf {R_{\mu}}\\
\mathbf {R_{\rho}}\\
\end{pmatrix}
\end{aligned} 
\end{equation}
, where $\mathbf {G_{{l2}_{\lambda}}}$,$\mathbf {G_{{l2}_{\mu}}}$,$\mathbf {G_{{l2}_{\rho}}}$ are the gradient for $\lambda$, $\mu$, ${\rho}$.   $\mathbf{R_{\lambda}}$,$\mathbf{R_{\mu}}$,$\mathbf{R_{\rho}}$ are the regulation terms and the mathematical expressions for these regulation terms are: 
\begin{equation}
\begin{aligned}
\begin{pmatrix}
\mathbf {R_{\lambda}} \\
\mathbf {R_{\mu}}\\
\mathbf {R_{\rho}}\\
\end{pmatrix} = 
\begin{pmatrix}
&{\alpha_{1}^{\lambda}}  \mathbf {\mathcal L^{T}_x} \mathbf { Q_{x_{\lambda}}}  \hspace{0.1in}
&{\alpha_{1}^{\lambda}}  \mathbf {\mathcal L^{T}_z} \mathbf { Q_{z_{\lambda}}},  \hspace{0.1in}
&{\alpha_{2}^{\lambda}}  \mathbf {\mathcal K^{T}_{xx}} \mathbf { Q_{{xx}_{\lambda}}}  \hspace{0.1in}
&{\alpha_{2}^{\lambda}}  \mathbf {\mathcal K^{T}_{zz}} \mathbf { Q_{{zz}_{\lambda}}} \\
&{\alpha_{1}^{\mu}}  \mathbf {\mathcal L^{T}_x} \mathbf { Q_{{x}_{\mu}}}   \hspace{0.1in}
&{\alpha_{1}^{\mu}}  \mathbf {\mathcal L^{T}_z}  \mathbf { Q_{{z}_{\mu}}}   \hspace{0.1in}
&{\alpha_{2}^{\mu}}  \mathbf {\mathcal K^{T}_{xx}} \mathbf { Q_{{xx}_{\mu}}}   \hspace{0.1in}
&{\alpha_{2}^{\mu}}  \mathbf {\mathcal K^{T}_{zz}} \mathbf { Q_{{zz}_{\mu}}}  \\
&{\alpha_{1}^{\rho}}  \mathbf {\mathcal L^{T}_x} \mathbf { Q_{{x}_{\rho}}}   \hspace{0.1in}
&{\alpha_{1}^{\rho}}  \mathbf {\mathcal L^{T}_z} \mathbf { Q_{{z}_{\rho}}}  \hspace{0.1in}
&{\alpha_{2}^{\rho}}  \mathbf {\mathcal K^{T}_{xx}} \mathbf { Q_{{xx}_{\rho}}}  \hspace{0.1in}
&{\alpha_{2}^{\rho}}  \mathbf {\mathcal K^{T}_{zz}} \mathbf { Q_{{zz}_{\rho}}} 
\end{pmatrix} 
\begin{pmatrix}
\mathbf {\mathcal L_x}\\
\mathbf {\mathcal L_z}\\
\mathbf {\mathcal K_{xx}}\\ 
\mathbf {\mathcal K_{zz}} 
\end{pmatrix}
\cdot
\begin{pmatrix}
\mathbf {m_{\lambda}}\\
\mathbf{m_{\mu}} \\
\mathbf{m_{\rho}}
\end{pmatrix}\\
\end{aligned}
\end{equation}

 \begin{equation}
  \begin{pmatrix}
\mathbf {q_{{x}_{\lambda}}}   & \mathbf {q_{{x}_{\mu}}}  & \mathbf {q_{{x}_{\rho}}} \\
\mathbf {q_{{z}_{\lambda}}}   & \mathbf {q_{{z}_{\mu}}}  & \mathbf {q_{{z}_{\rho}}}\\
\mathbf {q_{{xx}_{\lambda}}}   & \mathbf {q_{{xx}_{\mu}}}  & \mathbf {q_{{xx}_{\rho}}} \\
\mathbf {q_{{zz}_{\lambda}}}   & \mathbf {q_{{zz}_{\mu}}}  & \mathbf {q_{{zz}_{\rho}}}\\
 \end{pmatrix} = 
 \begin{pmatrix}
\mathbf  {\mathcal L_x}\\
\mathbf {\mathcal L_z} \\
\mathbf {\mathcal K_{xx}} \\
 \mathbf {\mathcal K_{zz}} \\
 \end{pmatrix}
  \begin{pmatrix}
 \mathbf  {m_{\lambda}}, 
 \mathbf {m_{\mu}} , 
 \mathbf {m_{\rho}} 
 \end{pmatrix}
\end{equation}

\begin{equation}
\begin{pmatrix}
\mathrm {Q_{{x}_{\lambda}}}   &   \mathrm {Q_{{x}_{\mu}}}         &\mathrm {Q_{{x}_{\rho}}}  \\
\mathrm {Q_{{z}_{\lambda}}}   &   \mathrm {Q_{{z}_{\mu}}}         &\mathrm {Q_{{z}_{\rho}}} \\
\mathrm {Q_{{xx}_{\lambda}}}   &   \mathrm {Q_{{xx}_{\mu}}}         &\mathrm {Q_{{xx}_{\rho}}}  \\
\mathrm {Q_{{zz}_{\lambda}}}   &   \mathrm {Q_{{zz}_{\mu}}}         &\mathrm {Q_{{zz}_{\rho}}}  \\
\end{pmatrix} = 
\begin{pmatrix}
\mathrm {\frac{1}{|q_{{x}_{\lambda}}|}}  & 
\mathrm {\frac{1}{|q_{{x}_{\mu}}|}}  & 
\mathrm {\frac{1}{|q_{{x}_{\rho}}|}} \\
\mathrm {\frac{1}{|q_{{z}_{\lambda}}|}} & 
\mathrm {\frac{1}{|q_{{z}_{\mu}}|}}   & 
\mathrm {\frac{1}{|q_{{z}_{\rho}}|}} \\
\mathrm {\frac{1}{|q_{{xx}_{\lambda}}|}}  & 
\mathrm {\frac{1}{|q_{{xx}_{\mu}}|}} & 
\mathrm {\frac{1}{|q_{{xx}_{\rho}}|}} \\
\mathrm {\frac{1}{|q_{{zz}_{\lambda}}|}} & 
\mathrm {\frac{1}{|q_{{zz}_{\mu}}|}} & 
\mathrm {\frac{1}{|q_{{zz}_{\rho}}|}} 
\end{pmatrix}
\end{equation} 

$\mathbf{T}$ means the transpose of the matrix, $\cdot$ meas dot product. 
$\mathbf{q_{{x}_{\mathbf {\lambda}}}}$ represent the first order TV regularization vector in x direction for parameter ${\lambda}$. $\mathrm {q_{{x}_{ {\lambda}}}}$ represent the values in vector $\mathbf{q_{{x}_{\lambda}}}$. 
$ \mathrm {Q_{{x}_{\lambda}}}$ is the absolute inverse of $\mathrm {q_{{x}_{\lambda}}}$. $\mathrm {Q_{{x}_{\lambda}}}$ are elements in vector $\mathbf{Q_{{x}_{\lambda}}}$. 
$\mathbf{q_{{xx}_{\mathbf {\lambda}}}}$ represent the second order TV regularization vector in x direction for parameter ${\lambda}$. $\mathrm {q_{{xx}_{ {\lambda}}}}$ represent the values in vector $\mathbf{q_{{xx}_{\lambda}}}$. 
$ \mathrm {Q_{{xx}_{\lambda}}}$ is the absolute inverse of $\mathrm {q_{{xx}_{\lambda}}}$. $\mathrm {Q_{{xx}_{\lambda}}}$ are elements in vector $\mathbf{Q_{{xx}_{\lambda}}}$. 
Other values in equations (9) and (10) can be also deduced like this. 
$\mathbf {\mathcal L_{x}}$, $\mathbf {\mathcal L_{z}}$ are the first order differential vector to give the first order total variations in x and z directions respectively. 
$\mathbf {\mathcal K_{xx}}$, $\mathbf {\mathcal K_{zz}}$ are the second order differential vector to give the second order total variations in x and z directions respectively. 

If we were to use $l1$ norm objective function with TV regulation. The objective function can be written as: 
\begin{equation}
\begin{aligned}
{\mathbf {\Phi}}_{l1}^{TV}
({\mathbf {m_{\lambda},{m_{\mu},m_{\rho}}}},
{\alpha_{1}^{\lambda}}, 
{\alpha_{1}^{\mu}}, 
{\alpha_{1}^{\rho}}, 
{\alpha_{2}^{\lambda}}, 
{\alpha_{2}^{\mu}}, 
{\alpha_{2}^{\rho}})  
=  
\lVert \mathbf {D_{syn}} \mathbf {(m_{\lambda},m_{\mu},m_{\rho})}-\mathbf {D_{obs}} \rVert  +\\
{\alpha_{1}^{\lambda}}  \mathbf {{\Theta}_{TV}} \mathbf {(m_{\lambda})}  +
{\alpha_{1}^{\mu}}  \mathbf {{\Theta}_{TV}} \mathbf {(m_{\mu})}  +
{\alpha_{1}^{\rho}}  \mathbf {{\Theta}_{{TV}}} \mathbf {(m_{\rho})}  + \\
{\alpha_{2}^{\lambda}}  \mathbf {\Upsilon_{TV}} \mathbf {(m_{\lambda})}  +
{\alpha_{2}^{\mu}}  \mathbf {\Upsilon_{TV}}  \mathbf {(m_{\mu})}  +
{\alpha_{2}^{\rho}}  \mathbf {\Upsilon_{TV}} \mathbf {(m_{\rho})} 
\end{aligned}
\end{equation}

\begin{figure}[h!]
	\centering
	\includegraphics[width=1\textwidth]{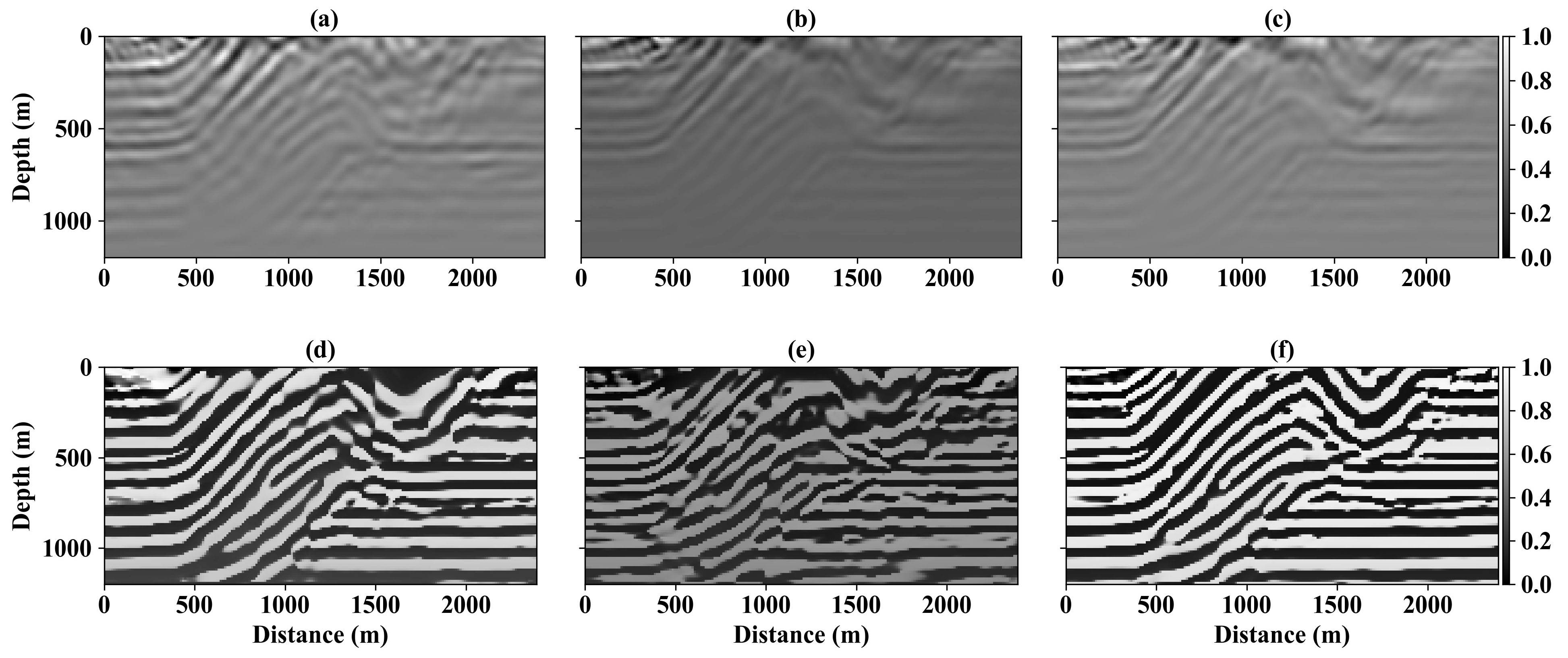}
	\caption{
		(a) ${\lambda}$ gradient given by the adjoint state method.
		(b) $\mu$ gradient based by the adjoint state method.
		(c) $\rho$ gradient based by the adjoint state method.
		(d) ${\lambda}$ gradient given by the AD method.
		(e) $\mu$ gradient based by the AD method.
		(f) $\rho$ gradient based by the AD method.
	}\label{fig:CFWI_ADFWI_GRAD.jpg}
\end{figure}

The gradient for each parameter based on l1 norm:
\begin{equation}
\begin{aligned}
\begin{pmatrix} 
\frac{{\partial}  {\mathbf {\Phi}}_{l1}^{TV} }{{\partial} {\mathbf{m_{\lambda}}}}\\
\frac{{\partial}  {\mathbf {\Phi}}_{l1}^{TV} }{{\partial} {\mathbf{m_{\mu}}}}\\
\frac{{\partial}  {\mathbf {\Phi}}_{l1}^{TV} }{{\partial} {\mathbf{m_{\rho}}}}\\
\end{pmatrix} = 
\begin{pmatrix}
\mathbf {G_{{l1}_{\lambda}}} \\
\mathbf {G_{{l1}_{\mu}}} \\
\mathbf {G_{{l1}_{\rho}}} \\
\end{pmatrix}+
\begin{pmatrix}
\mathbf {R_{\lambda}} \\
\mathbf {R_{\mu}}\\
\mathbf {R_{\rho}}\\
\end{pmatrix}
\end{aligned} 
\end{equation}
, where $\mathbf {G_{{l1}_{\lambda}}}$,$\mathbf {G_{{l1}_{\mu}}}$,$\mathbf {G_{{l1}_{\rho}}}$ are the gradient for $\lambda$, $\mu$, ${\rho}$ using $l_1$ norm as misfit function. The gradient calculation in $l_1$ norm is using a differenta adjoint source (\cite{pyun2009frequency} \cite{brossier2010data}).  The adjoint source for the adjoint fields for $l_1$ norm is , In the case of real arithmetic numbers, the term $\frac{{\Delta \mathbf{d} }}{| \Delta \mathbf{d} |}$  corresponds to the function $sign$. In this study, we did not meet conditions when $\Delta \mathbf{d} = 0$. The detail gradient expression using the adjoint state method for parameters $\lambda$, $\mu$ and $\rho$ based on $l_2$ and $l_1$ norm can be expressed as:
\begin{equation}
\begin{array}{l}
\mathbf {G_{{l2}_{{\lambda}}}}=  -
\sum_{\mathbf{x_{s}}} \sum_{\mathbf{x_{g}}} 
\int_{0}^{T} \\
\left(\left(\partial_{x} \tilde{u}_{x}\left(\mathbf{r}, \mathbf{r}_{s}, t\right)+\partial_{z} \tilde{u}_{z}\left(\mathbf{r}, \mathbf{r}_{s}, t\right)\right) 
\left(\partial_{x} \tilde{u}_{x}^{*_{l2}}\left(\mathbf{r}, \mathbf{r}_{g}, T-t\right)+\partial_{z} \tilde{u}_{z}^{*_{l2}}\left(\mathbf{r}, \mathbf{r}_{g}, T-t\right)\right)\right)
\end{array}
\end{equation}

\begin{equation}
\begin{array}{l}
\mathbf {G_{{l2}_{\mu}}}=  -
\sum_{\mathbf{x_{s}}} \sum_{\mathbf{x_{g}}} 
\int_{0}^{T} \\
\left(\left(\partial_{z} \tilde{u}_{x}\left(\mathbf{r}, \mathbf{r}_{s}, t\right)+
\partial_{x} \tilde{u}_{z}\left(\mathbf{r}, \mathbf{r}_{s}, t\right)\right)\left
(\partial_{z} \tilde{u}_{x}^{*_{l2}}\left(\mathbf{r}, \mathbf{r}_{g}, T-t\right)+\partial_{x} \tilde{u}_{z}^{*_{l2}}\left(\mathbf{r}, \mathbf{r}_{g}, T-t\right)\right)\right) \\ -2\left(\left(\partial_{x} \tilde{u}_{x}\left(\mathbf{r}, \mathbf{r}_{s}, t\right) \partial_{x} \tilde{u}_{x}^{*_{l2}}\left(\mathbf{r}, \mathbf{r}_{g}, T-t\right)+
\partial_{z} \tilde{u}_{z}\left(\mathbf{r}, \mathbf{r}_{s}, t\right) \partial_{z} \tilde{u}_{z}^{*_{l2}}\left(\mathbf{r}, \mathbf{r}_{g}, T-t\right)\right)\right)
\end{array}
\end{equation}

\begin{equation}
\mathbf {G_{{l2}_{\rho}}} = 
\sum_{\mathbf{x_{s}}} \sum_{\mathbf{x_{g}}} 
\int_{0}^{T}  \\
\left(\left(
\partial_{t} \tilde{u}_{x}\left(\mathbf{r}, \mathbf{r}_{s}, t\right) 
\partial_{t} \tilde{u}_{x}^{*_{l2}}\left(\mathbf{r}, \mathbf{r}_{g}, T-t\right)+
\partial_{t} \tilde{u}_{z}\left(\mathbf{r}, \mathbf{r}_{s}, t\right) 
\partial_{t} \tilde{u}_{z}^{*_{l2}}\left(\mathbf{r}, \mathbf{r}_{g}, T-t\right)
\right)\right)
\end{equation}

\begin{equation}
\begin{array}{l}
\mathbf {G_{{l1}_{{\lambda}}}}=  -
\sum_{\mathbf{x_{s}}} \sum_{\mathbf{x_{g}}} 
\int_{0}^{T} \\
\left(\left(\partial_{x} \tilde{u}_{x}\left(\mathbf{r}, \mathbf{r}_{s}, t\right)+\partial_{z} \tilde{u}_{z}\left(\mathbf{r}, \mathbf{r}_{s}, t\right)\right)\left(\partial_{x} \tilde{u}_{x}^{*_{l1}}\left(\mathbf{r}, \mathbf{r}_{g}, T-t\right)+\partial_{z} \tilde{u}_{z}^{*_{l1}}\left(\mathbf{r}, \mathbf{r}_{g}, T-t\right)\right)\right)
\end{array}
\end{equation}

\begin{equation}
\begin{array}{l}
\mathbf {G_{{l1}_{\mu}}}=  -
\sum_{\mathbf{x_{s}}} \sum_{\mathbf{x_{g}}} 
\int_{0}^{T}  \\
\left(\left(\partial_{z} \tilde{u}_{x}\left(\mathbf{r}, \mathbf{r}_{s}, t\right)+
\partial_{x} \tilde{u}_{z}\left(\mathbf{r}, \mathbf{r}_{s}, t\right)\right)\left
(\partial_{z} \tilde{u}_{x}^{*_{l1}}\left(\mathbf{r}, \mathbf{r}_{g}, T-t\right)+\partial_{x} \tilde{u}_{z}^{*_{l1}}\left(\mathbf{r}, \mathbf{r}_{g}, T-t\right)\right)\right) \\ -2\left(\left(\partial_{x} \tilde{u}_{x}\left(\mathbf{r}, \mathbf{r}_{s}, t\right) \partial_{x} \tilde{u}_{x}^{*_{l1}}\left(\mathbf{r}, \mathbf{r}_{g}, T-t\right)+
\partial_{z} \tilde{u}_{z}\left(\mathbf{r}, \mathbf{r}_{s}, t\right) \partial_{z} \tilde{u}_{z}^{*_{l1}}\left(\mathbf{r}, \mathbf{r}_{g}, T-t\right)\right)\right)
\end{array}
\end{equation}

\begin{equation}
\mathbf {G_{{l1}_{\rho}}} = \\
\sum_{\mathbf{x_{s}}} \sum_{\mathbf{x_{g}}} 
\int_{0}^{T} 
\left(\left(\partial_{t} \tilde{u}_{x}\left(\mathbf{r}, \mathbf{r}_{s}, t\right) \partial_{t}\tilde{u}_{x}^{*_{l1}}\left(\mathbf{r}, \mathbf{r}_{g}, T-t\right)+
\partial_{t}\tilde{u}_{z}\left(\mathbf{r}, \mathbf{r}_{s}, t\right) \partial_{t} \tilde{u}_{z}^{*_{l1}}\left(\mathbf{r}, \mathbf{r}_{g}, T-t\right)\right)\right)
\end{equation}
$\mathbf {G_{{l2}_{\lambda}}} $,
$\mathbf {G_{{l2}_{\mu}}} $,
$\mathbf {G_{{l2}_{\rho}}} $,
are gradients for $\lambda$, $\mu$ and $\rho$ using $l_2$ norm as misfit respectively. 
$\mathbf {G_{{l1}_{\lambda}}} $,
$\mathbf {G_{{l1}_{\mu}}} $,
$\mathbf {G_{{l1}_{\rho}}} $,
are gradients for $\lambda$, $\mu$ and $\rho$ using $l_1$ norm as misfit respectively. $\tilde{u_{x}}^{*_{l1}}$  and $\tilde{u_{z}}^{*_{l1}}$  are the adjoint wavefields  generated by the l1 norm adjoint source,  $\tilde{u_{x}}^{*_{l2}}$  and $\tilde{u_{z}}^{*_{l2}}$  are the adjoint wavefields  generated by the l2 norm adjoint source. $T$ is the total receiving time for the shot records. $\mathbf {r_{s}}$,  $\mathbf {r_{g}}$ represent the source and receivers locations respectively.  $\mathbf {r}$ represent the model perturbation locations for $\lambda$, $\mu$ and $\rho$ model. Figure \ref{fig:CFWI_ADFWI_GRAD.jpg} shows the gradient calculated using the adjoint state method and the Automatic Difference method. Figure \ref{fig:CFWI_ADFWI_GRAD.jpg}  (a), (b), (c) are the normalized ${\lambda}$, ${\mu}$ and $\rho$ gradients calculated by using the adjont state method. Figure \ref{fig:CFWI_ADFWI_GRAD.jpg}  (d), (e), (f) are the normalized  ${\lambda}$, ${\mu}$ and $\rho$ gradients calculated by using the Automatic Differential method. The gradients calculated by using the Automatic Difference method, contains more information about the model, for instance the lower part of the model,  indicating that they can better reconstruct the mode. 
Now we rewrite the misfit function as:
\begin{equation}
\begin{aligned}
{\mathbf {\Phi}}^{TV}  
= \mathbf {J_{D}} +\mathbf {J_{r1}} + \mathbf {J_{r2}} 
\end{aligned}
\end{equation}
,where 
$\mathbf {J_{D}}$ represents the any kind of norm misfit between observed data and synthetic data.
$\mathbf {J_{r1}}  = {\alpha_{1}^{\lambda}}  \mathbf {{\Theta}_{TV}} \mathbf {(m_{\lambda})}  +
{\alpha_{1}^{\mu}}  \mathbf {{\Theta}_{TV}} \mathbf {(m_{\mu})}  +
{\alpha_{1}^{\rho}}  \mathbf {{\Theta}_{{TV}}} \mathbf {(m_{\rho})} $. 
$\mathbf {J_{r2}}  = {\alpha_{2}^{\lambda}}  \mathbf {\Upsilon_{TV}} \mathbf {(m_{\lambda})}  +
{\alpha_{2}^{\mu}}  \mathbf {\Upsilon_{TV}}  \mathbf {(m_{\mu})}  +
{\alpha_{2}^{\rho}}  \mathbf {\Upsilon_{TV}} \mathbf {(m_{\rho})} $. 
The value $\alpha_{1}^{\lambda}$, ${\alpha_{1}^{\mu}} $, ${\alpha_{1}^{\rho}}$, $\alpha_{2}^{\lambda}$, 
${\alpha_{2}^{\mu}}$, 
${\alpha_{2}^{\rho}}$. 
are chosen according to the following formula (\cite{guitton2012constrained}). 
\begin{equation}
T = \frac{\mathbf {J_{D}}}{\mathbf {J_{r1}} +\mathbf {J_{r2}} }
\end{equation}
 We should control the  values for $\alpha_{1}^{\lambda}$, ${\alpha_{1}^{\mu}} $, ${\alpha_{1}^{\rho}}$, $\alpha_{2}^{\lambda}$, 
${\alpha_{2}^{\mu}}$, 
${\alpha_{2}^{\rho}}$ and keep value T between 1 and 10. T should be relatively large when, noise occurs in the data (\cite{xiang2016efficient}).

\subsection{Parameterization testing}

By modifying the RNN cells, and changing the trainable parameters, we can examine the influence of parameterization on waveform inversion within the deep learning formulation.  Three sets of parameter classes are considered: the velocity parameterization (D-V model), involving P-wave velocity, S-wave velocity, and density; the modulus parameterization (D-M model), involving the Lam\'{e} parameters ${\lambda}$ and ${\mu}$, and density; and, the stiffness matrix model (D-S model), involving ${C_{11}}$, $C_{44}$ and density ${\rho}$.

\begin{figure}[h!]
	\centering
	\includegraphics[width=0.6\textwidth]{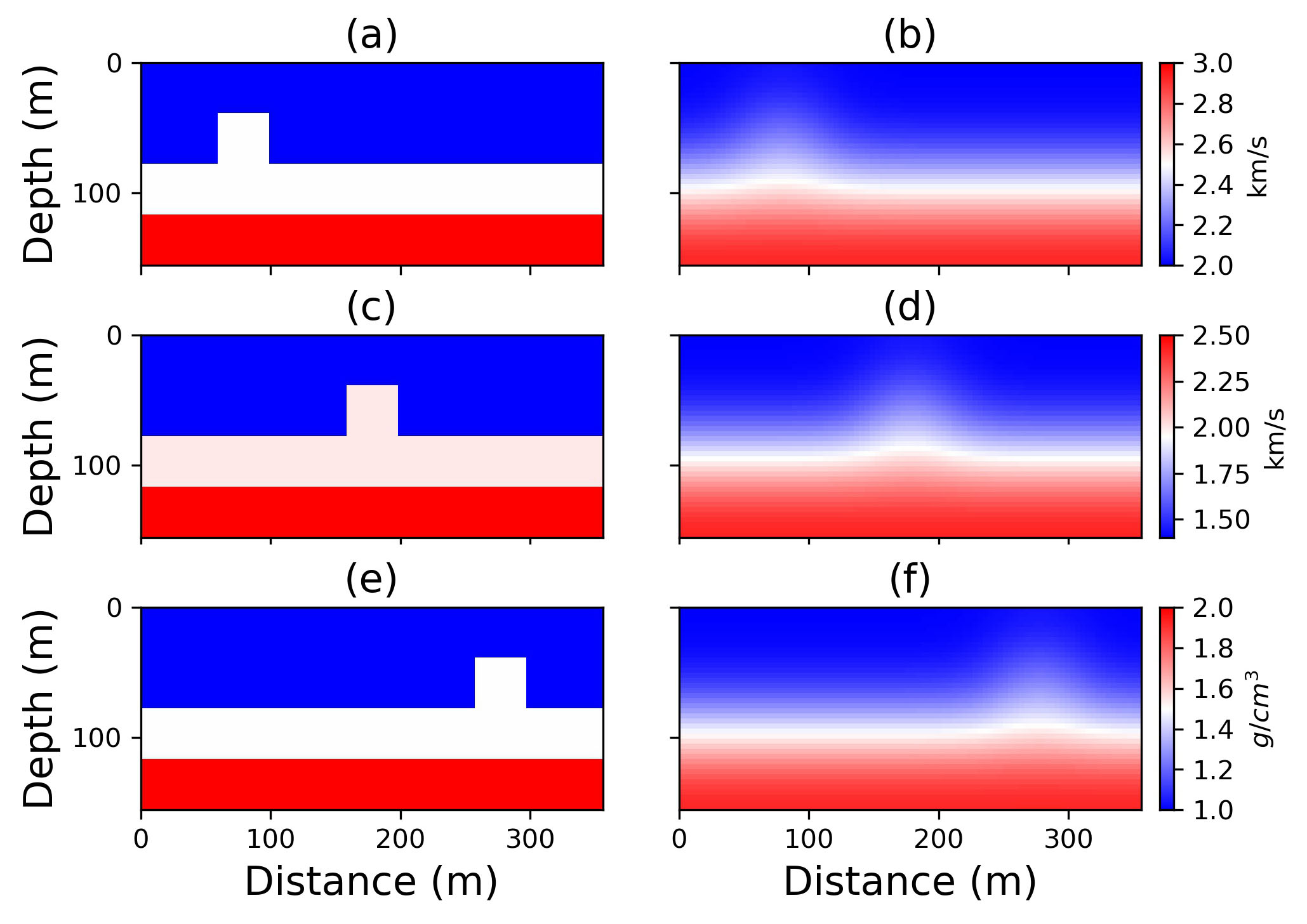}
	\caption{(a) true Vp model, (b) initial Vs model, (c) true Vs model, (d) initial vs model, (e) true ${\rho }$ mode, (f) initial ${\rho}$ model.  }\label{fig:0903_vp_vs_rho_layers.jpg}
\end{figure}

 In these tests, the size of each model is 40${\times}90$. 7 source points are evenly distributed across the surface of the model; the source is a Ricker wavelet with a dominant frequency of $30Hz$. The grid length of the model is $dx=dz=4$m. In Figure \ref{fig:0903_vp_vs_rho_layers.jpg} (a)-(b) are true and initial $V_P$ model, (c)-(d) are the true and initial $V_S$ model, (e)-(f)  are the true and initial $\rho$ model we use in this test.

\begin{figure}[h!]
\centering
\includegraphics[width=1\textwidth]{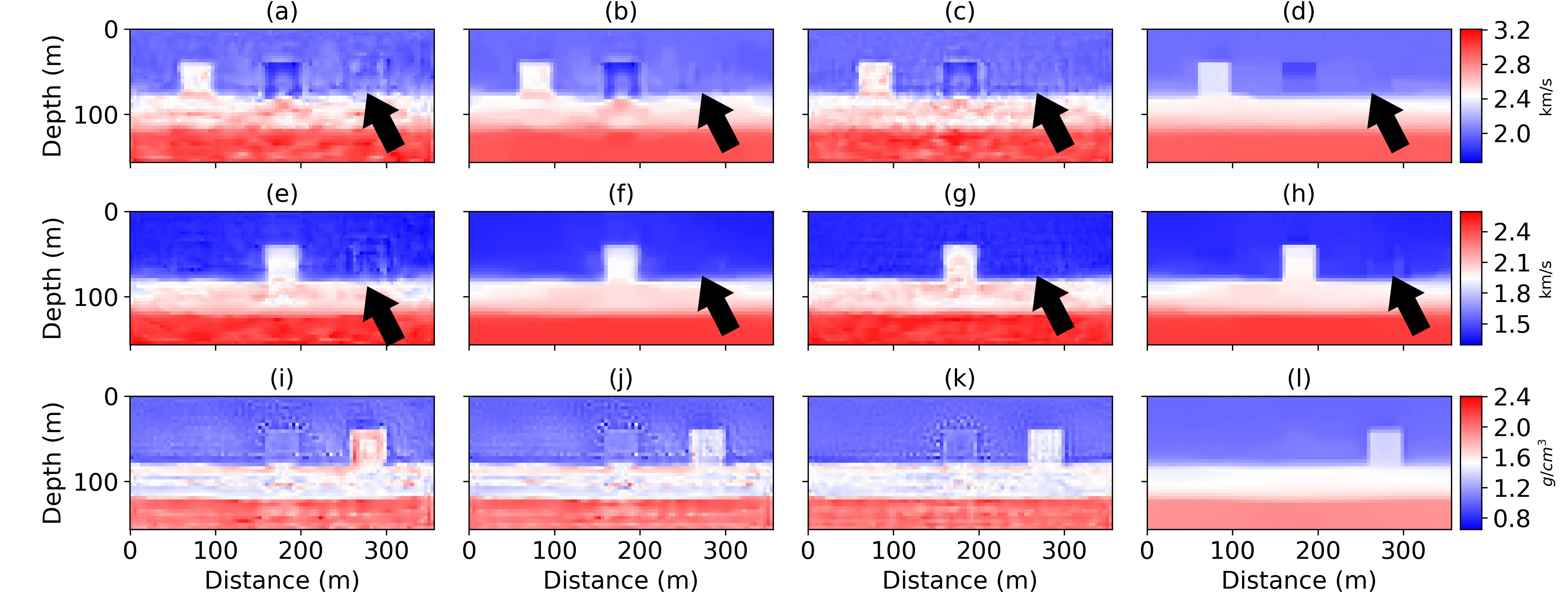}
\caption{D-V parameterization inversion results. (a)-(d), $V_P$ $l_2$ norm,  $l_2^{TV}$ norm, $l_1$ norm, $l_1^{TV}$ norm inversion results respectively,  (e)-(h), $V_S$ $l_2$ norm,  $l_2^{TV}$ norm, $l1$ norm $l_1^{TV}$ norm,  inversion results respectively,  (i)-(l), $\rho$ $l_2$ norm $l_2^{TV}$ norm, $l_1$ norm,  $l_1^{TV}$ norm inversion results respectively.  }\label{fig:VD_layers.jpg}
\end{figure}

In Figures \ref{fig:VD_layers.jpg} are the inversion results using the D-V parameterization. Figure \ref{fig:VD_layers.jpg} (a)-(d) are the inversion results for $V_P$ generated by $l_2$,  $l_2^{TV}$, $l_1$ and $l_1^{TV}$ norm.  Figure \ref{fig:VD_layers.jpg} (e)-(h) are the inversion results for $V_S$ generated by $l_2$,  $l_2^{TV}$, $l_1$ and $l_1^{TV}$ norm.  Figure \ref{fig:VD_layers.jpg} (i)-(l) are the inversion results for $\rho$ generated by $l_2$,  $l_2^{TV}$, $l_1$ and $l_1^{TV}$ norm.  In Figure \ref{fig:VD_layers.jpg}  (a), (e) and (i) we can see the cross talk between different parameters as the back arrows indicate, In Figure \ref{fig:VD_layers.jpg}  (b), (c), (d) we can see that by changing the misfits and adding high order TV regulations, the cross talk between $V_P$ and $\rho$ has been reduced.  In Figure \ref{fig:VD_layers.jpg}  (h) and (l) we can see that by using $l_1^{TV}$ norm, the cross talk between density and $V_S$ has been mitigated, while in (j) and (k) we still see the cross talk between $V_S$ and density. From  Figures \ref{fig:VD_layers.jpg}  we can see that $l_1$ norm with high order TV regulation can help to mitigate the cross talk problem. 

\begin{figure}[h!]
\centering
\includegraphics[width=1\textwidth]{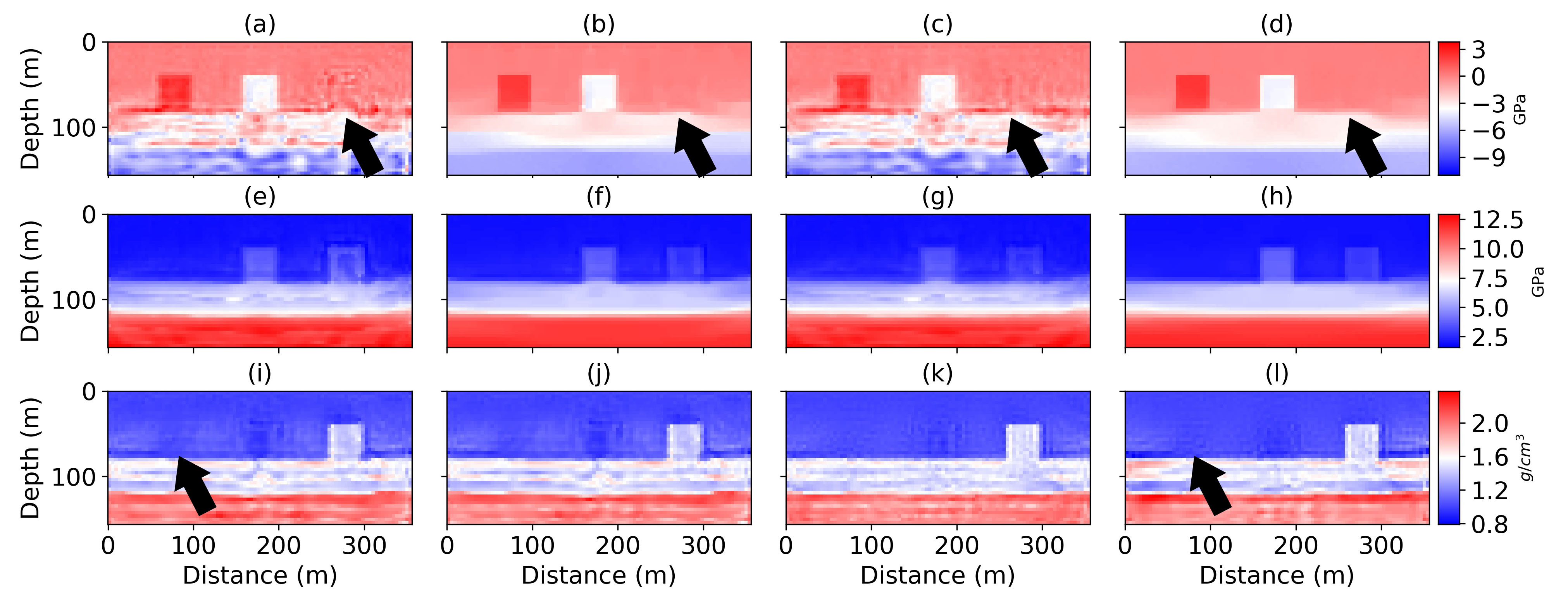}
\caption{D-M parameterization inversion results. (a)-(d), ${\lambda}$ $l_2$, norm, $l_2^{TV}$ norm, $l_1$, norm $l_1^{TV}$ norm inversion results respectively,  (e)-(h), $\mu$ $l_2$ norm, $l_2^{TV}$ norm, $l1$ norm, $l_1^{TV}$ norm inversion results respectively,  (i)-(l), $\rho$ $l_2$ norm, $l_2^{TV}$ norm, $l_1$ norm, $l_1^{TV}$ norm inversion results respectively.  }\label{fig:MD_layers.jpg}
\end{figure}

Next the modulus parameterization is examined, in which we seek to recover ${\lambda}$, ${\mu}$ and ${\rho}$ models.  This occurs through a straightforward modification of the RNN cell, and again a change in the trainable parameters from velocities to moduli. Figure \ref{fig:MD_layers.jpg} (a)-(d) are the inversion results for $\lambda$ generated by $l_2$,  $l_2^{TV}$, $l_1$ and $l_1^{TV}$ norm.  Figure \ref{fig:MD_layers.jpg} (e)-(h) are the inversion results for $\mu$ generated by $l_2$,  $l_2^{TV}$, $l_1$ and $l_1^{TV}$ norm.  Figure \ref{fig:MD_layers.jpg} (i)-(l) are the inversion results for $\rho$ generated by $l_2$,  $l_2^{TV}$, $l_1$ and $l_1^{TV}$ norm.  In Figure \ref{fig:MD_layers.jpg} (b) and (d) we can see that by using the high order TV regulations in $l_2$ and $l_1$ norm the cross talk between density and $\lambda$ has been mitigated. Figure (j) shows that In this parameterization by using the  high order TV regulation on $l_2$ norm can provide better inversion results for density as well. 

\begin{figure}[h!]
\centering
\includegraphics[width=1\textwidth]{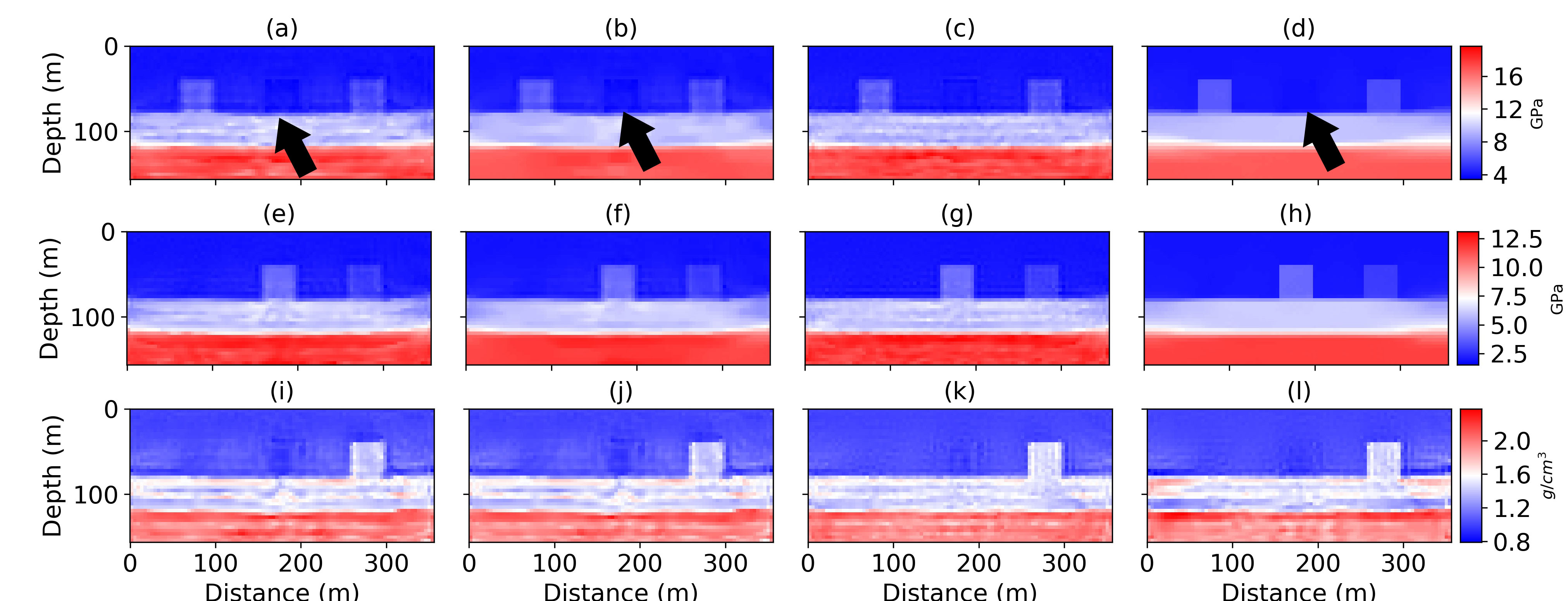}
\caption{D-S parameterization inversion results. (a)-(d), $c11$ $l_2$ norm,  $l_2^{TV}$ norm, $l_1$ norm, $l_1^{TV}$ norm inversion results respectively,  (e)-(h), $c44$ $l_2$ norm,  $l_2^{TV}$ norm, $l1$ norm $l_1^{TV}$ norm,  inversion results respectively,  (i)-(l), $\rho$ $l_2$ norm $l_2^{TV}$ norm, $l_1$ norm,  $l_1^{TV}$ norm inversion results respectively. }\label{fig:SD_layers.jpg}
\end{figure}

Inversion results for models in the S-D parameterization are plotted in Figure \ref{fig:SD_layers.jpg}, generated using the change of variables ${C_{11}} = V_P^{2}{\rho}$ and ${C_{44}} = V_S^{2}{\rho}$.  Figure \ref{fig:SD_layers.jpg} (a)-(d) are the inversion results for $\lambda$ generated by $l_2$,  $l_2^{TV}$, $l_1$ and $l_1^{TV}$ norm.  Figure \ref{fig:SD_layers.jpg} (e)-(h) are the inversion results for $\mu$ generated by $l_2$,  $l_2^{TV}$, $l_1$ and $l_1^{TV}$ norm.  Figure \ref{fig:SD_layers.jpg} (i)-(l) are the inversion results for $\rho$ generated by $l_2$,  $l_2^{TV}$, $l_1$ and $l_1^{TV}$ norm. In Figure \ref{fig:SD_layers.jpg} we can still see the cross talk between $c44$ and $c11$ as the black arrows pointing out. However this cross talk has been mitigated by in figure (d), by using the $l_1^{TV}$ norm misfit.  The inversion results above shows that the RNN based high order TV regulation FWi based on the $l_2$ and $l_1$ norm has the ability to mitigate cross talk problem with only gradient based methods. The RNN based $l_2^{TV}$ RNN FWI in D-M parameterization and $l_1^{TV}$ RNN FWI in D-S parameterization provide better inversion results than other inversion tests. 

\begin{figure}[h!]
	\centering
	\includegraphics[width=0.7\textwidth]{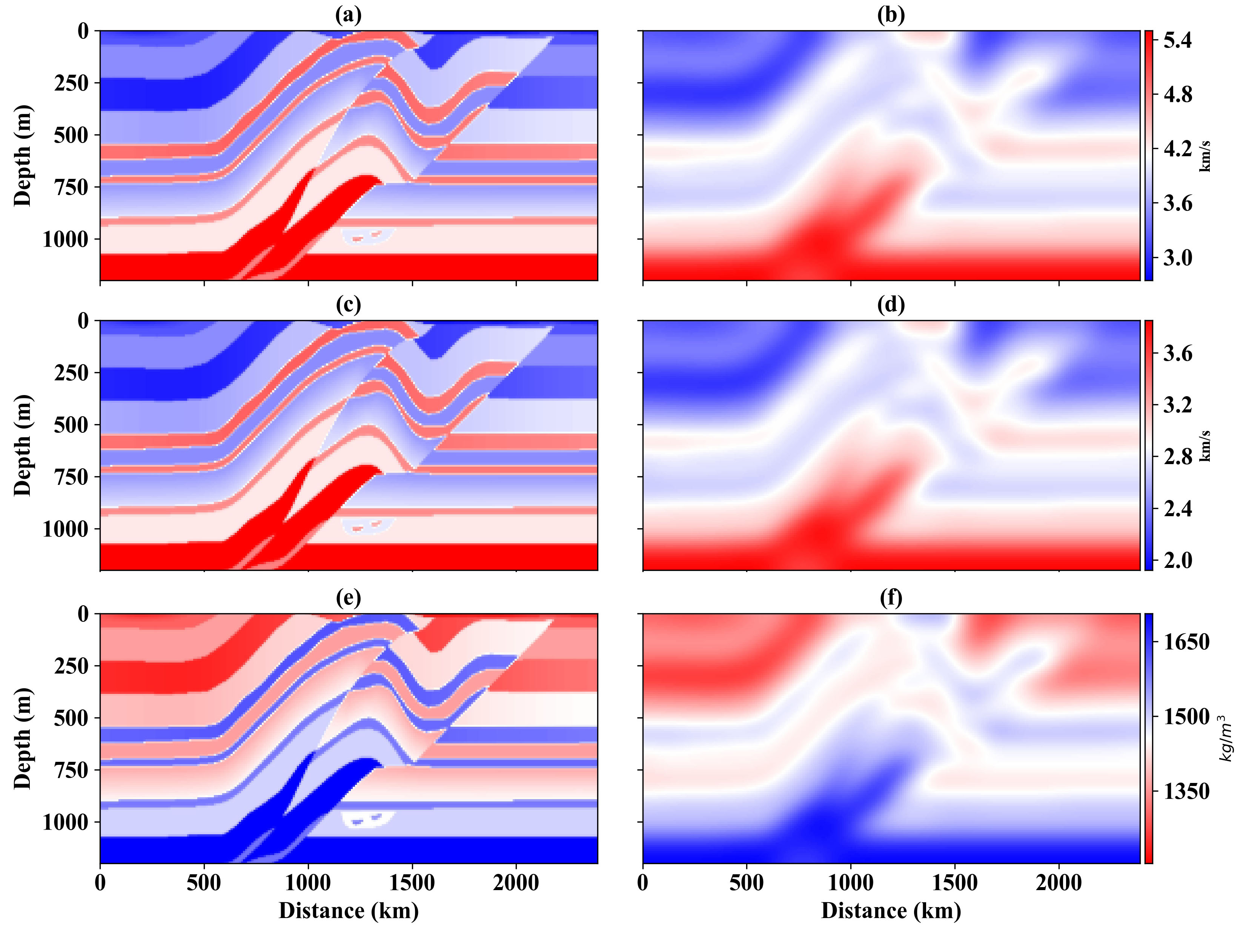}
	\caption{(a) true $V_P$ model, (b) initial $V_P$ model, 
		(c) true  $V_S$ model (d) initial Vs model,  
		(e) ${\rho}$ true (f) initial $\rho$}
	\label{fig:Over_true_vp_vs_rho.jpg}
\end{figure}

\begin{figure}[h!]
	\centering
	\includegraphics[width=0.4\textwidth]{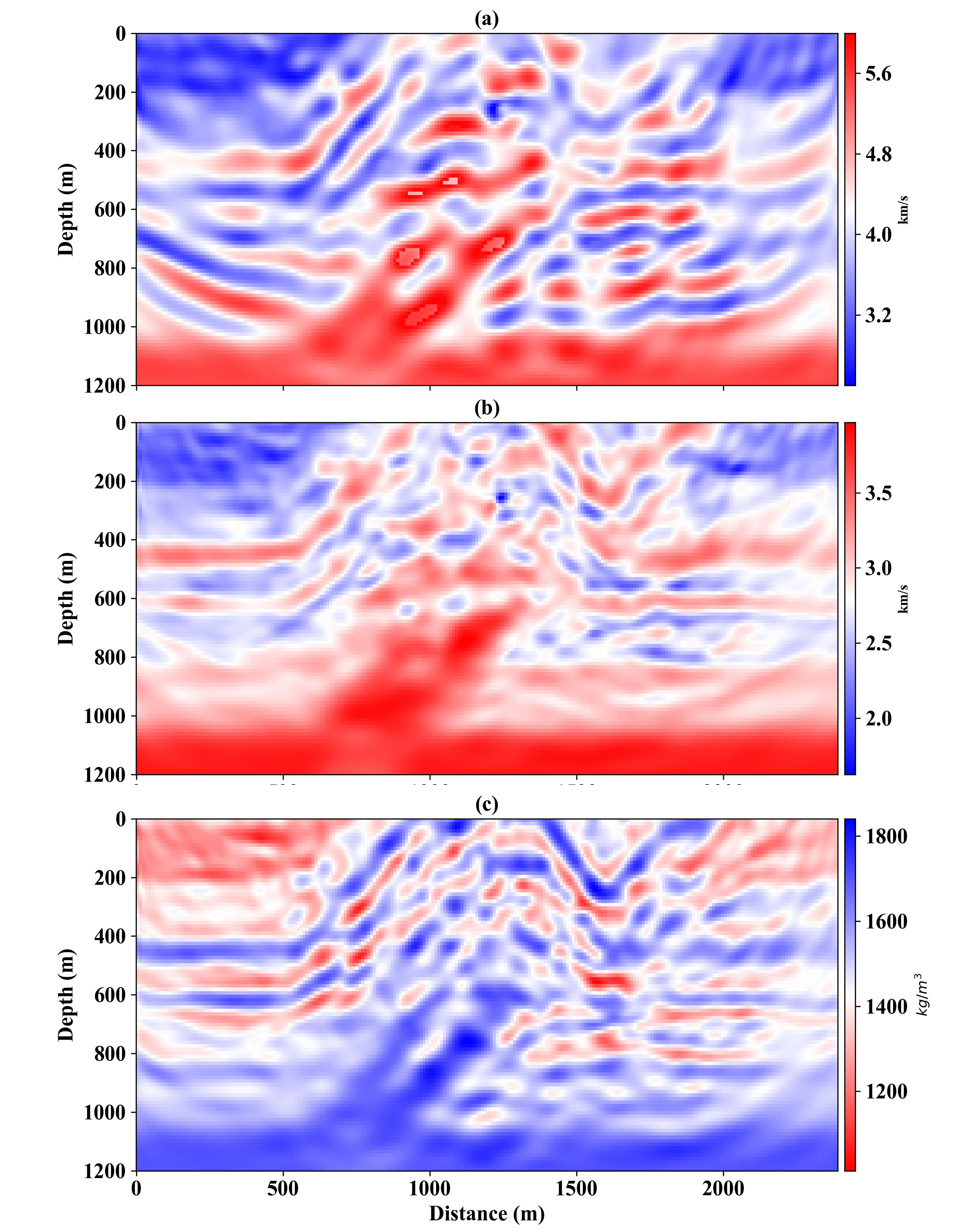}
	\caption{(a) true $V_P$ model, (b) initial $V_P$ model, 
		(c) true  $V_S$ model (d) initial $V_S$ model,  
		(e) ${\rho}$ true (f) initial $\rho$}
	\label{fig:CFWI_results.jpg}
\end{figure}

\begin{figure}[h!]
	\centering
	\includegraphics[width=1\textwidth]{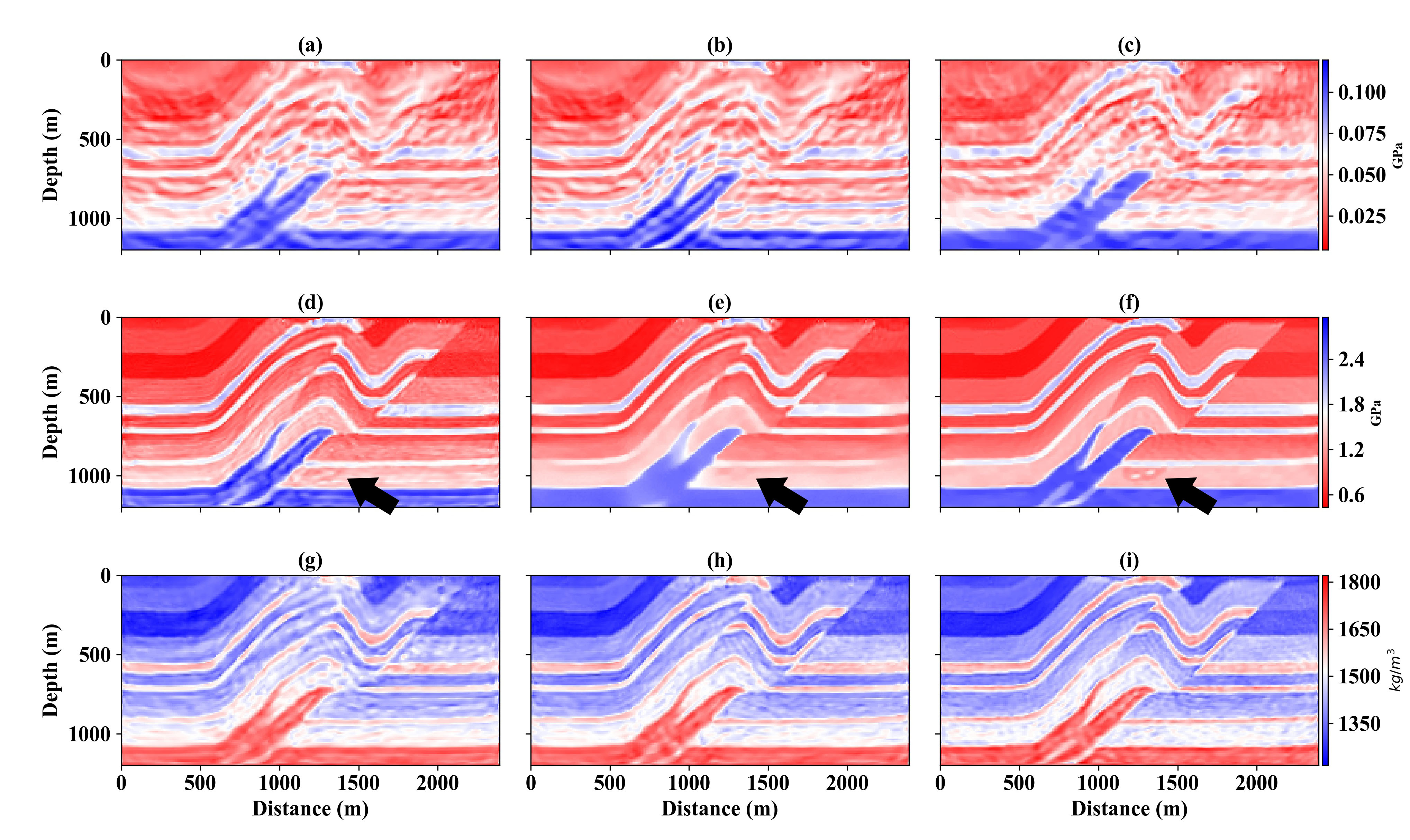}
	\caption{M-D parameterization inversion results. (a)-(c), $\lambda$ $l_2$ norm,  $l_2^{TV}$ norm, $l_1^{TV}$ norm inversion results respectively,  (c)-(f), $\mu$ $l_2$ norm,  $l_2^{TV}$ norm, $l_1^{TV}$ norm,  inversion results respectively,  (g)-(i), $\rho$ $l_2$ norm $l_2^{TV}$ norm, $l_1^{TV}$ norm inversion results respectively}\label{fig:Over_MD_lam_muu_rho_0903.jpg}
\end{figure}

\begin{figure}[h!]
	\centering
	\includegraphics[width=1\textwidth]{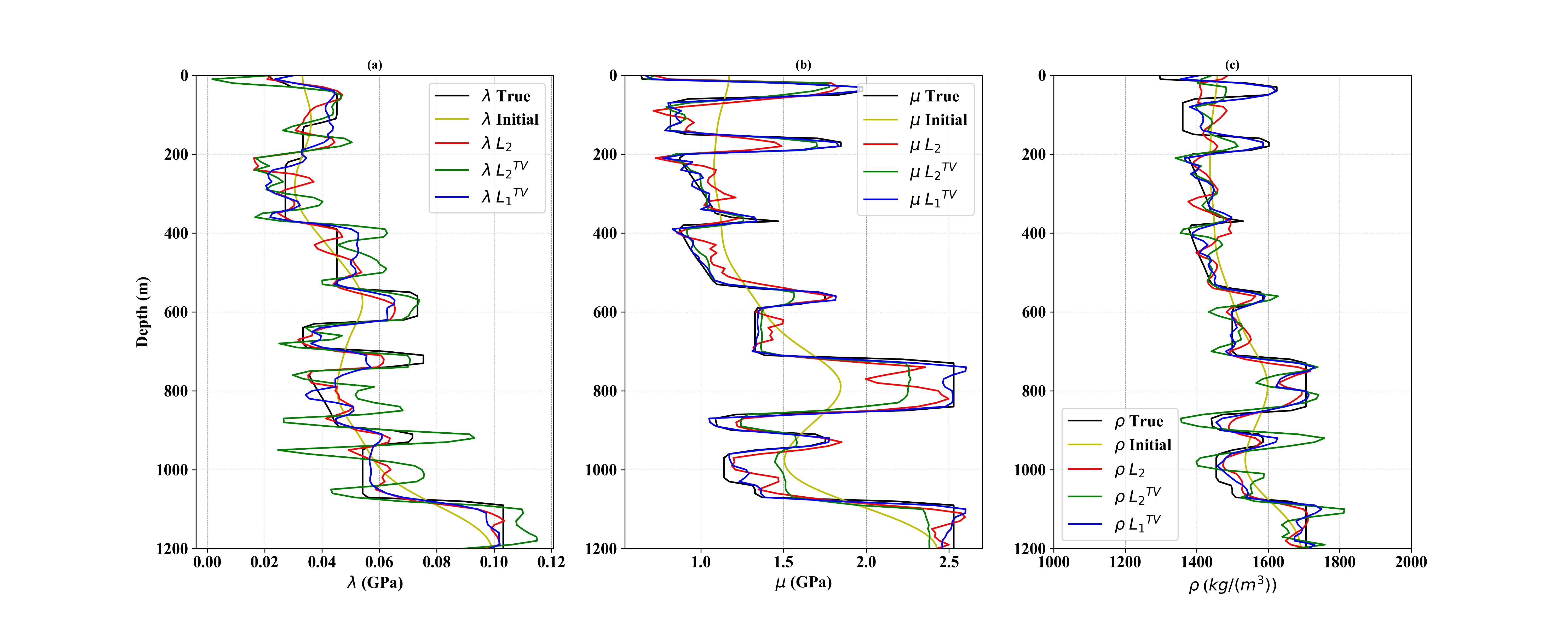}
	\caption{Profiles through the recovered elastic models.  (a) V\underline{}ertical $\lambda$ profiles; (b) vertical $\mu$ profiles; (c) vertical ${\rho}$ profiles}\label{fig:Over_MD_0903_profile.jpg}
\end{figure}

Next, we will verify the proposed methods on the over-thrust model. Figure \ref{fig:Over_true_vp_vs_rho.jpg} (a), (c), (e) demonstrate the true models for $V_P$, $V_S$, and density $\rho$ model, and (b), (d) and (h) are the initial models for $V_P$, $V_S$, and density $\rho$ respectively. The size of the model is 121 $\times$ 240. The grid length of the model is 10m. 12 shots are evenly distributed on the surface of the model and every grid point has a receiver. The source of the wavelet is Ricker's wavelet with main frequency 20Hz.

Figure Figure \ref{fig:CFWI_results.jpg}  shows the inversion results by using the conventional FWI. Figure Figure \ref{fig:CFWI_results.jpg} (a) is the inversion for $V_P$, Figure Figure \ref{fig:CFWI_results.jpg} (b) is the inversion for $V_S$, Figure Figure \ref{fig:CFWI_results.jpg} (c) is the inversion for $\rho$. From figure \ref{fig:CFWI_results.jpg} we can see that the overall inversion resolution by using the conventional FWI is poor. 

Figure \ref{fig:Over_MD_lam_muu_rho_0903.jpg}  shows the inversion results by using D-M parameterization. Figure \ref{fig:Over_MD_lam_muu_rho_0903.jpg} (a)-(c) are ${\lambda}$ $l_2$ norm,  $l_2^{TV}$ norm, $l_1^{TV}$ norm inversion results respectively,  (d)-(f) are $\mu$ $l_2$ norm,  $l_2^{TV}$ norm, $l_1^{TV}$ norm,  inversion results respectively,  (g)-(i) are $\rho$ $l_2$ norm $l_2^{TV}$ norm, $l_1^{TV}$ norm inversion results respectively.  In this parameterization we get unstable inversion results for parameter $\lambda$. 
However,  in D-M parameterization Figure \ref{fig:Over_MD_0903_profile.jpg} (f), the small half arc structure at around 1000m of the model has been recovered in , as the black arrows indicate. Figure \ref{fig:Over_MD_0903_profile.jpg} shows the profiles through the recovered elastic modules at 1000m of the models based on D-M parameterization. In  Figure \ref{fig:Over_MD_0903_profile.jpg}, the black lines are the true values, the yellow lines are the initial values, red lines are the inversion results for $l_2$ norm, green lines are the inversion results for $l_2^{TV}$ norm and blue lines are the inversion results for $l_1^{TV}$ norm. Compared with the true lines we can also see that , Figure \ref{fig:Over_MD_0903_profile.jpg}  (b) and  Figure \ref{fig:Over_MD_0903_profile.jpg}  (c), $l_1^{TV}$ norm inversion results are more close to the true values.

\begin{figure}[h!]
	\centering
	\includegraphics[width=1\textwidth]{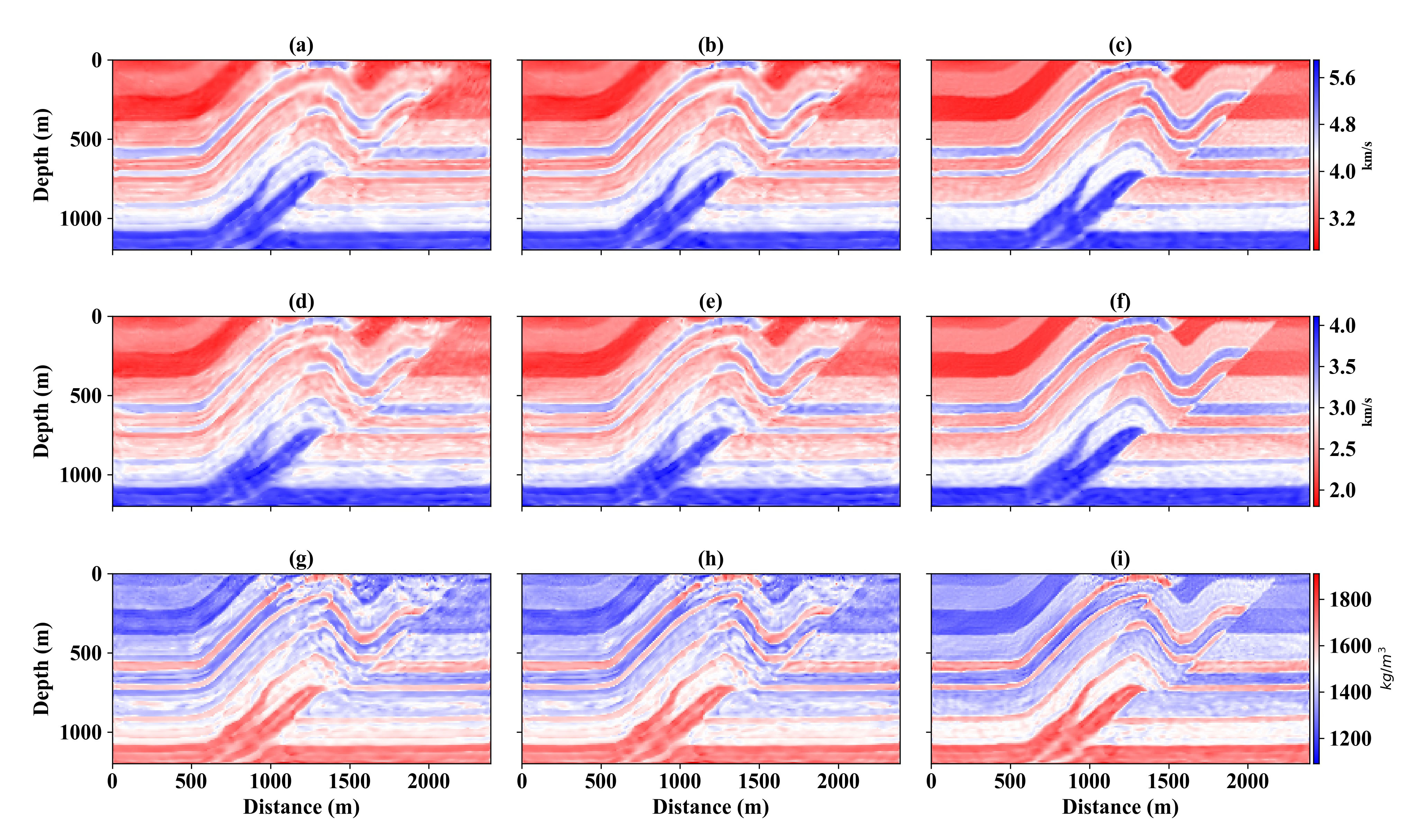}
	\caption{D-V parameterization inversion results. (a)-(d), $V_P$ $l_2$ norm,  $l_2^{TV}$ norm, $l_1^{TV}$ norm inversion results respectively,  (e)-(h), $V_S$ $l_2$ norm,  $l_2^{TV}$ norm, $l_1^{TV}$ norm,  inversion results respectively,  (i)-(l), $\rho$ $l_2$ norm $l_2^{TV}$ norm, $l_1^{TV}$ norm inversion results respectively.}\label{fig:Over_VD_vp_vs_rho_0903.jpg}
\end{figure}

\begin{figure}[h!]
	\centering
	\includegraphics[width=1\textwidth]{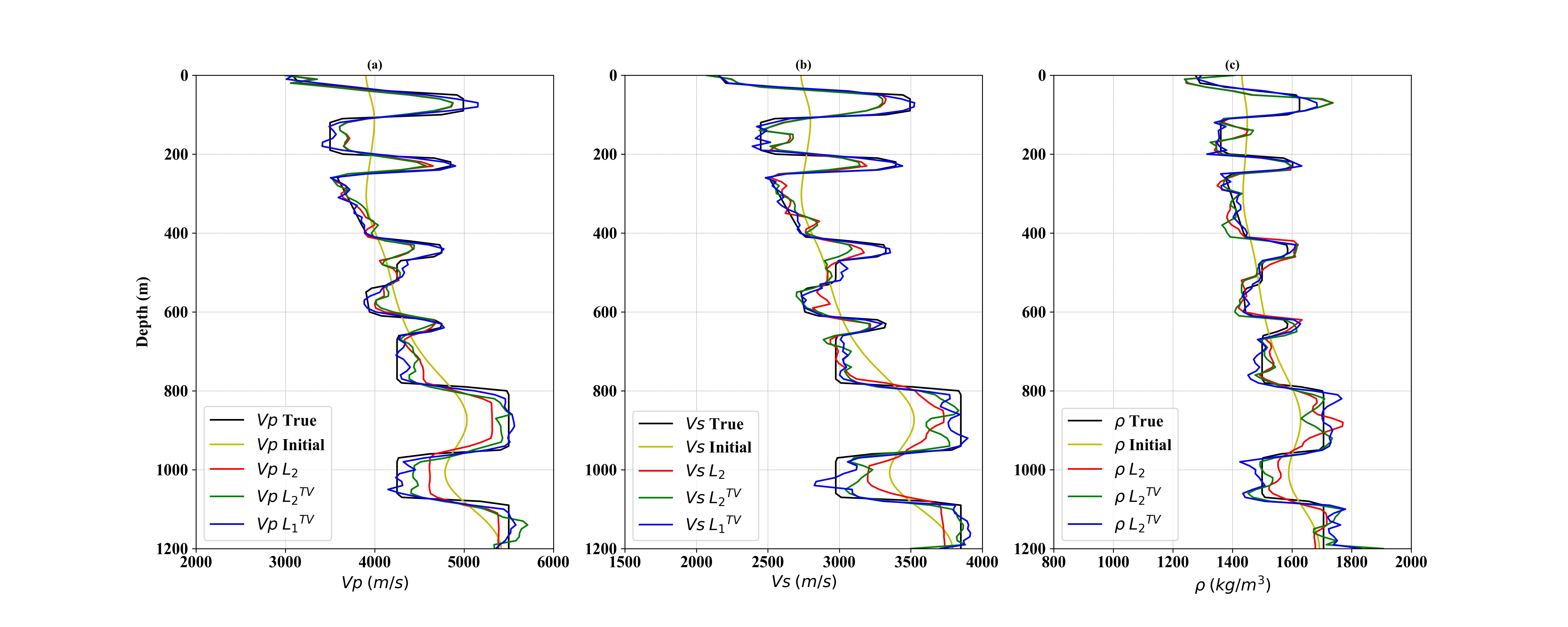}
	\caption{Profiles through the recovered elastic models.  (a) V\underline{}ertical $V_P$ profiles; (b) vertical $V_S$ profiles; (c) vertical ${\rho}$ profiles.}\label{fig:Over_VD_0903_profile.jpg}
\end{figure}

\begin{figure}[h!]
	\centering
	\includegraphics[width=0.9\textwidth]{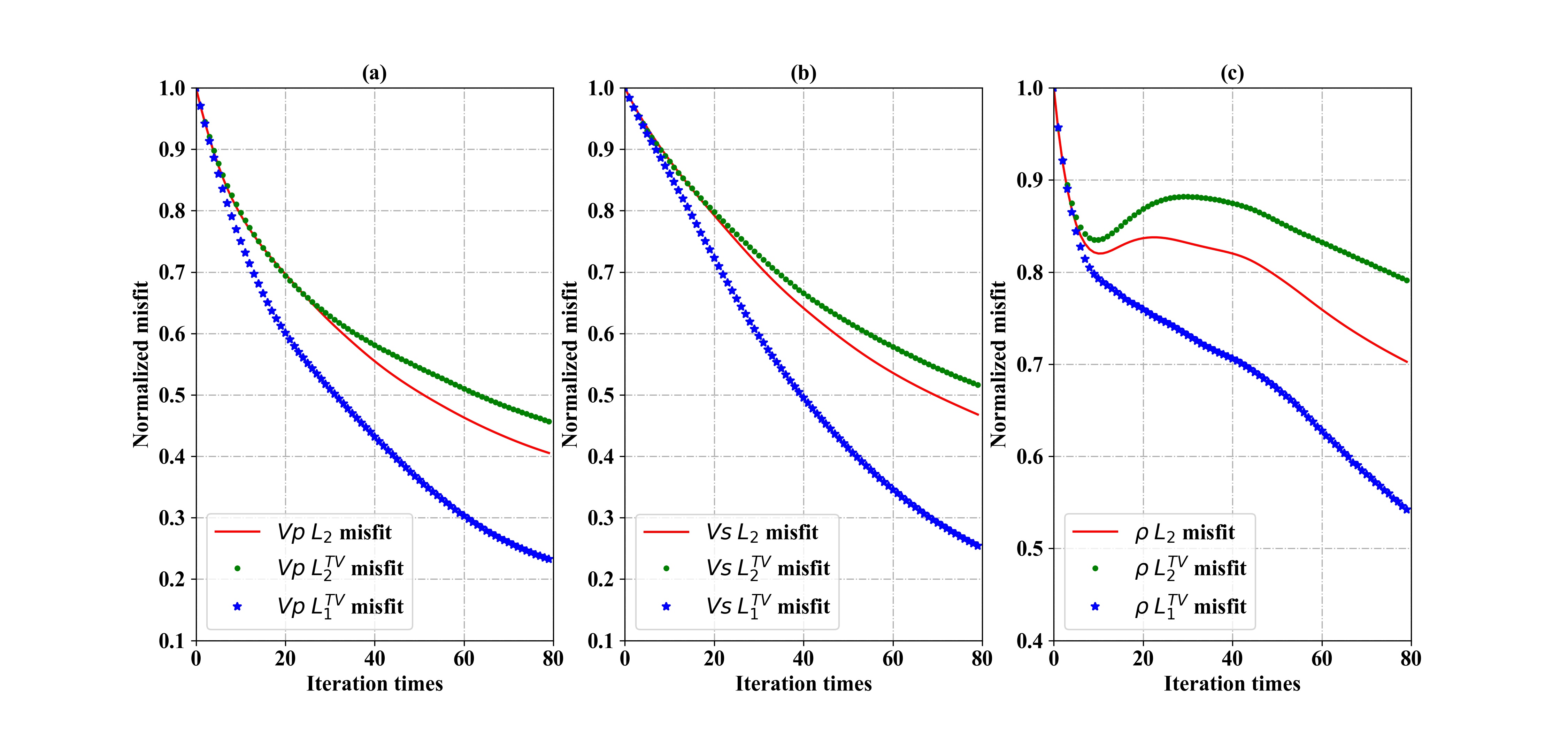}
	\caption{D-V parameterization inversion results . (a) Vp model misfit using $l_2$,$l_2^{TV}$ and $l_1^{TV}$ norm.(b) Vs model misfit using $l_2$,$l_2^{TV}$ and $l_1^{TV}$ norm.(c) $\rho$ model misfit using $l_2$,$l_2^{TV}$ and $l_1^{TV}$ norm }\label{fig:Over_VD_0903_model_loss.jpg}
\end{figure}

Figure \ref{fig:Over_VD_vp_vs_rho_0903.jpg}  shows the inversion results by using D-V parameterization. In this parameterization all the three parameters are stable.  Figure \ref{fig:Over_VD_vp_vs_rho_0903.jpg} (a)-(c) are, $V_P$ $l_2$ norm,  $l_2^{TV}$ norm, $l_1^{TV}$ norm inversion results respectively,  (d)-(f) are $V_S$ $l_2$ norm,  $l_2^{TV}$ norm, $l_1^{TV}$ norm,  inversion results respectively,  (g)-(i) are $\rho$ $l_2$ norm $l_2^{TV}$ norm, $l_1^{TV}$ norm inversion results respectively. Compared with the true models ,we can see that the inversion results generated by using the $l_2$ norm are less robust compared with other misfits. $l_1$ norm with high order TV regulation can provide more accurate inversion results. This can also be seen from Figure \ref{fig:Over_VD_vp_vs_rho_0903.jpg}, which is the profiles through the recovered elastic modules at 1000m of the models. In  Figure \ref{fig:Over_VD_vp_vs_rho_0903.jpg}, the black lines are the true velocities and density, the yellow lines are the initial values, red lines are the inversion results for $l_2$ norm, green lines are the inversion results for $l_2^{TV}$ norm
and blue lines are the inversion results for $l_1^{TV}$ norm.  Figure \ref{fig:Over_VD_0903_profile.jpg} (a) shows the results for $V_P$. Figure \ref{fig:Over_VD_0903_profile.jpg} (b) shows the results for $V_S$. Figure \ref{fig:Over_VD_0903_profile.jpg} (c) shows the results for $\rho$. In all the three figures we can see that blue lines are closer to the true values compared with other lines, which means that the $l_1^{TV}$ norm can provide us with more accurate inversion results.  Figure \ref{fig:Over_VD_0903_model_loss.jpg} shows the D-V parameterization inversion model misfits in each iteration. The red lines are the inversion using $l_2$ norm.  The green lines are the inversion using $l_2^{TV}$ norm.  The red lines are the inversion using $l_1^{TV}$ norm. Figure \ref{fig:Over_VD_0903_model_loss.jpg} shows that we can get higher accuracy inversion results by using the $l_1^{TV}$ norm with fewer iterations.

\begin{figure}[h!]
	\centering
	\includegraphics[width=1\textwidth]{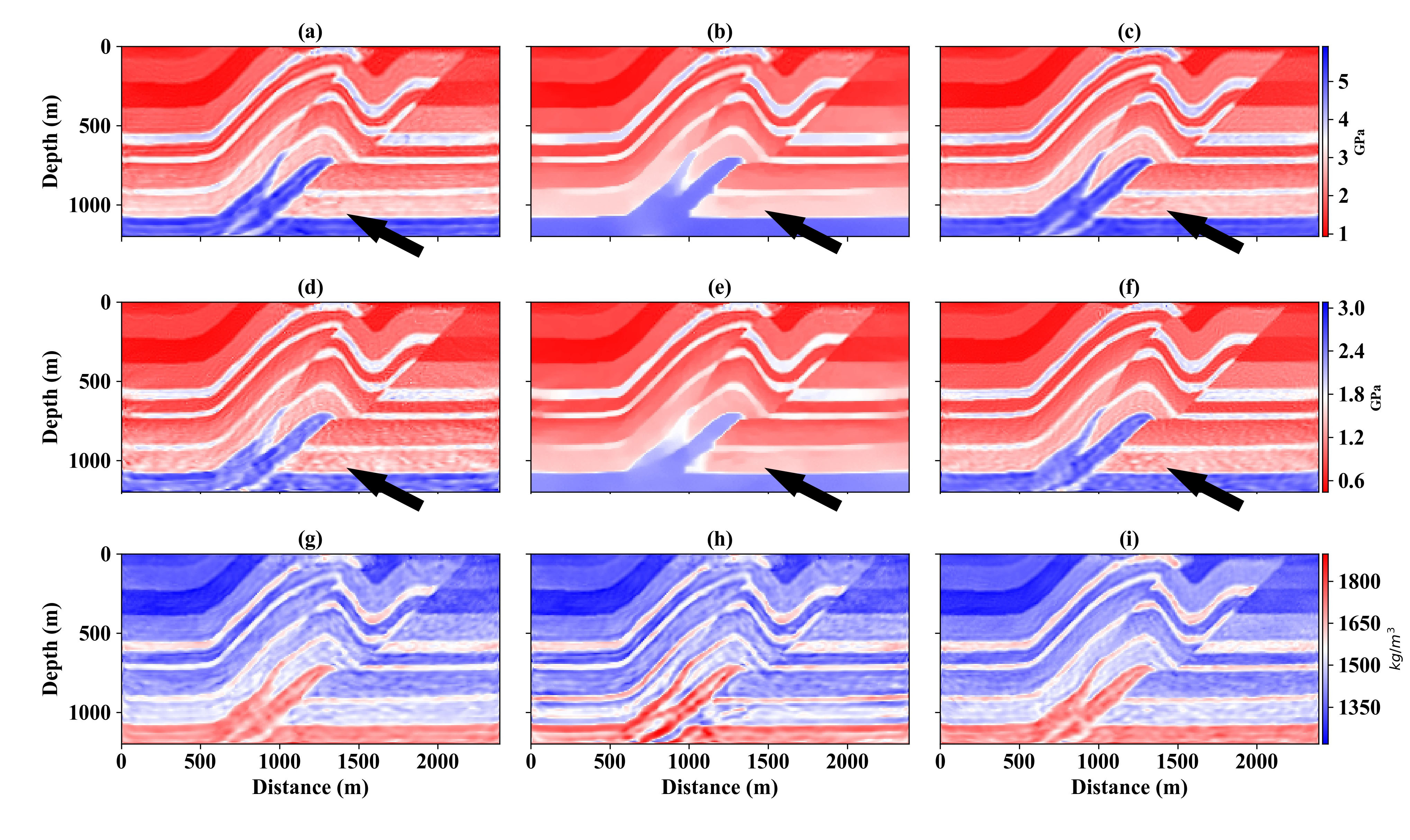}
	\caption{S-D parameterization inversion results. (a)-(c), $c11$ $l_2$ norm,  $l_2^{TV}$ norm, $l_1^{TV}$ norm inversion results respectively,  (c)-(f), $c44$ $l_2$ norm,  $l_2^{TV}$ norm, $l_1^{TV}$ norm,  inversion results respectively,  (g)-(i), $\rho$ $l_2$ norm $l_2^{TV}$ norm, $l_1^{TV}$ norm inversion results respectively.}\label{fig:Over_SD_c11_c44_rho_0903.jpg}
\end{figure}

\begin{figure}[h!]
	\centering
	\includegraphics[width=1\textwidth]{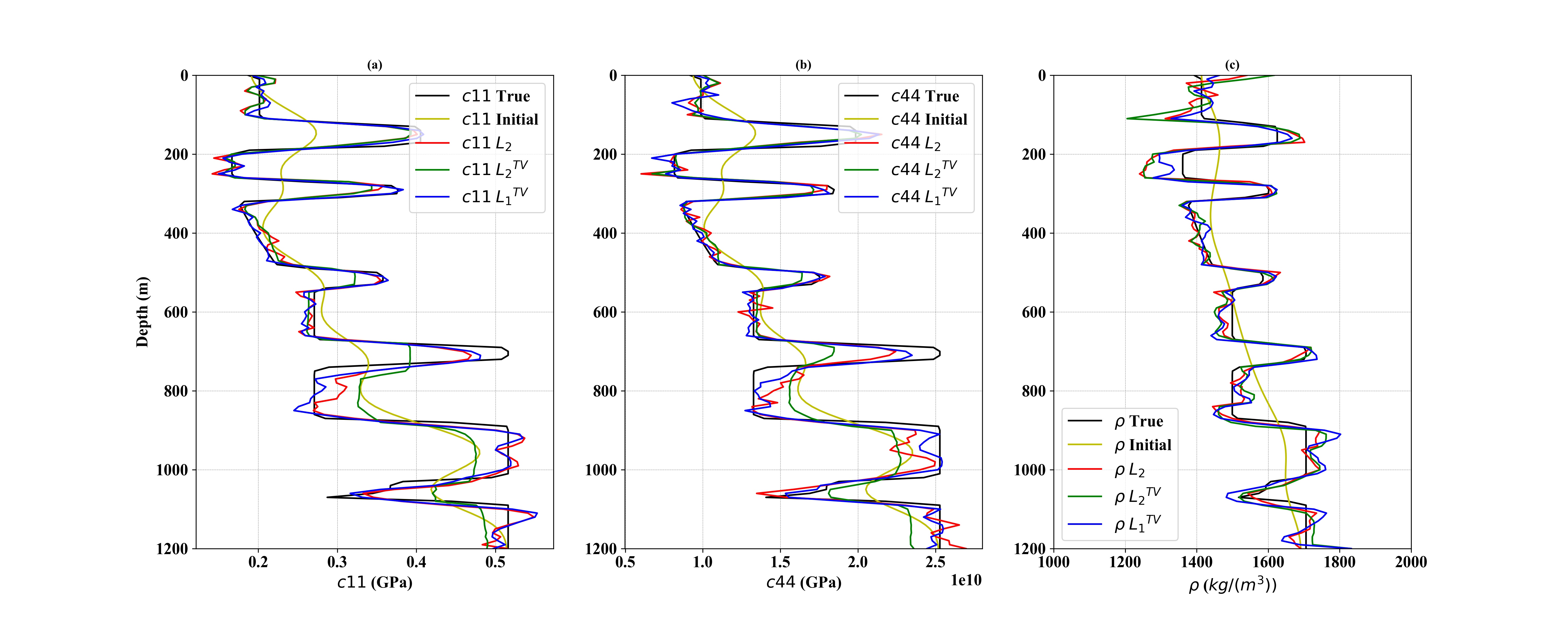}
	\caption{Profiles through the recovered elastic models based on .  (a) V\underline{}ertical $c11$ profiles; (b) vertical $c44$ profiles; (c) vertical ${\rho}$ profiles}\label{fig:Over_SD_0903_profile.jpg}
\end{figure}

Figure \ref{fig:Over_SD_c11_c44_rho_0903.jpg}  shows the inversion results by using D-S parameterization. Figure \ref{fig:Over_SD_c11_c44_rho_0903.jpg} (a)-(c),  are ${c11}$ $l_2$ norm,  $l_2^{TV}$ norm, $l_1^{TV}$ norm inversion results respectively,  (d)-(f) are $c44$ $l_2$ norm,  $l_2^{TV}$ norm, $l_1^{TV}$ norm,  inversion results respectively,  (g)-(i) are $\rho$ $l_2$ norm $l_2^{TV}$ norm, $l_1^{TV}$ norm inversion results respectively. In this parameterization the small half arc structure at around 1000m of the model is clearly resolved in this parameterization by using the $l_1^{TV}$ norm.  Compared with other misfits , again the $l_1^{TV}$ norm provide us the best resolution for the model. Figure \ref{fig:Over_MD_0903_profile.jpg} shows the profiles through the recovered elastic modules at 1000m of the models based on D-M parameterization.. Red lines, green lines and blue lines are $l_2$, $l_2^{TV}$, and $l_1^{TV}$ norm inversion results respectively. We can also see that in D-S parameteriation. $l_1^{TV}$ provide us with inversion results that is more close the to true values.
Figure \ref{fig:Over_SD_0903_profile.jpg}  shows how model misfits in D-M parameterization changes in each iteration. The blue line, the $l_1^{TV}$ norm inversion, has the fastest model misfit decline rate. We can conclude in this parameterization $l_1^{TV}$ can generate more accuracy inversion results with fewer iterations.

\begin{figure}[h!]
	\centering
	\includegraphics[width=0.9\textwidth]{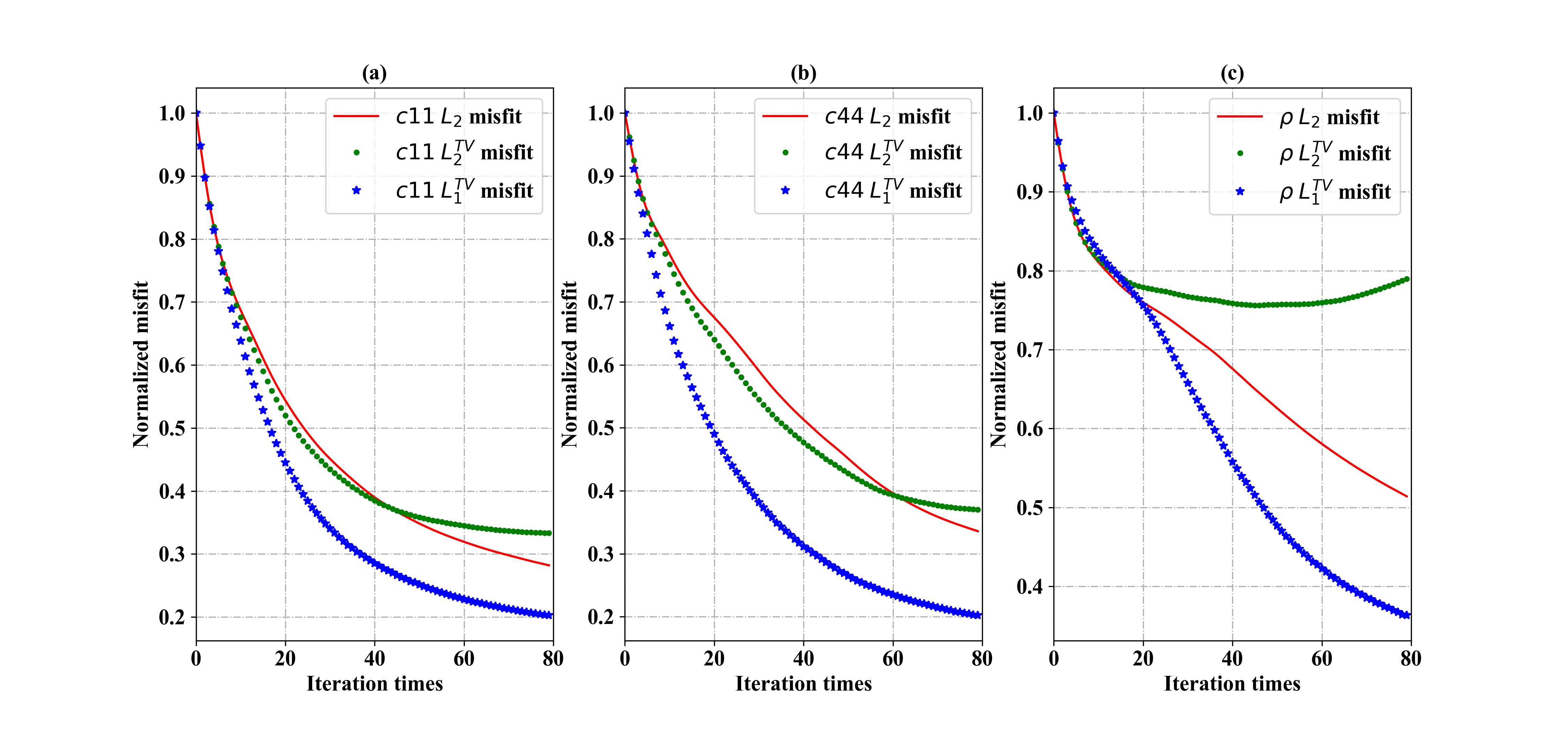}
	\caption{D-S parameterization inversion results . (a) $c11$ model misfit using $l_2$,$l_2^{TV}$ and $l_1^{TV}$ norm.(b) $c44$ model misfit using $l_2$,$l_2^{TV}$ and $l_1^{TV}$ norm.(c) $\rho$ model misfit using $l_2$,$l_2^{TV}$ and $l_1^{TV}$ norm}\label{fig:Fig_9_Cmatrix_layers_1110.jpg}
\end{figure}

\section{Random noise testing}
     
\begin{figure}[h!]
\centering
\includegraphics[width=1\textwidth]{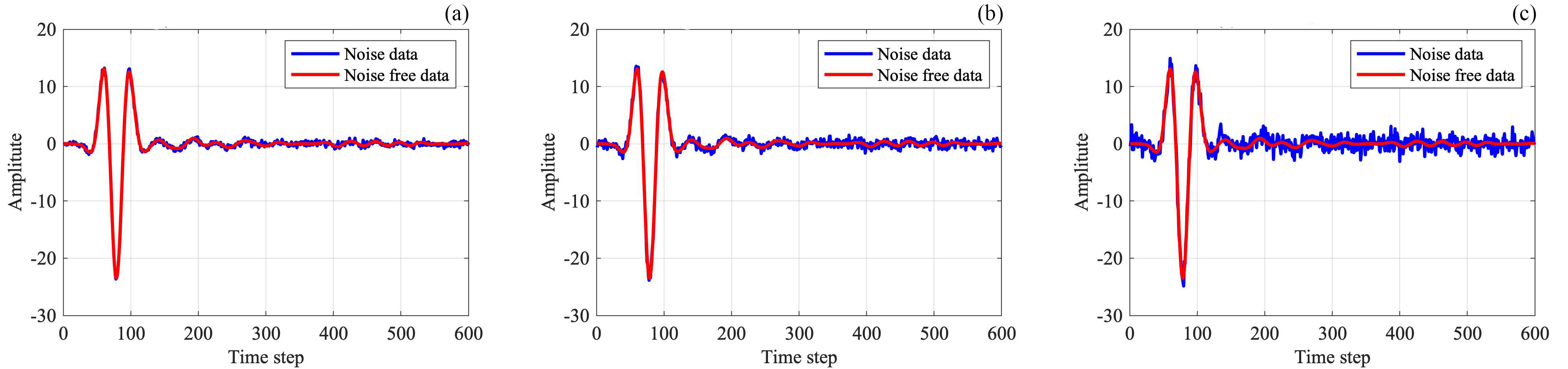}
\caption{Noise free and noise shotrecords. (a) N\underline{}oise free record and records with Gaussian noise $std$ = $0.3$.  (b) Noise free record and records with Gaussian noise $std$ = $0.5$. (c) Noise free record and records with Gaussian noise $std$ = $1$.}\label{fig:noise_records.jpg}
\end{figure}

In this section, we will test the sensitivity of this deep learning method to data contaminated with random noise drawn from a Gaussian distribution. In Figure \ref{fig:noise_records.jpg}, we plot a trace from an example shot profile with different ratios of noise. 
In (a), the red signal is the record without noise, and the blue line is the record with Gaussian random noise.  The mean value of the noise is zero, and the standard deviation is 0.3 ($std$ = $0.3$); in (b) the noise-free data and data with noise at $std=0.5$ are plotted; in (c) the noise-free data and the data with noise at $std=1$ are plotted. 

\begin{figure}[h!]
\centering
\includegraphics[width=1\textwidth]{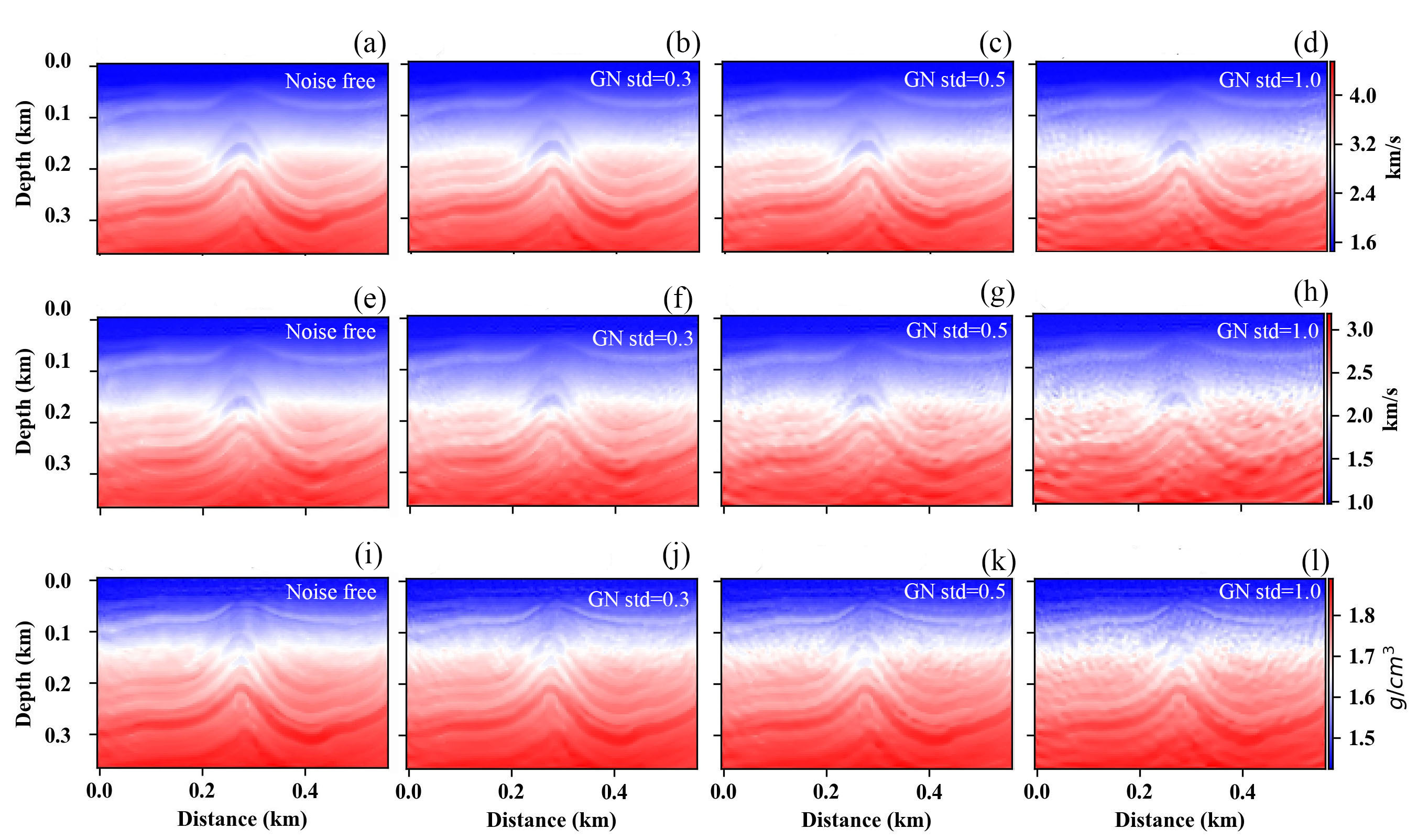}
\caption{Random noise testing inversion results. (a),(e),(i), noise free inversion results for $V_P$, $V_S$, and density (b),(f),(j) Gaussian noise $std=0.3$ inversion results for $V_P$, $V_S$, and density, (c),(g),(k) Gaussian noise $std=0.5$ inversion results for $V_P$, $V_S$, and density, (d),(h),(l) Gaussian noise $std=1.0$ inversion results for $V_P$, $V_S$, and density. }\label{fig:Noise_results.jpg}
\end{figure}

In Figure \ref{fig:Noise_results.jpg}a, e and i the inversion results from noise-free inversion based on the RNN are plotted.  In Figure \ref{fig:Noise_results.jpg}b, f and j, the inversion results with noise at $std=0.3$ for $V_P$, $V_S$, and ${\rho}$ are plotted, respectively; in Figure \ref{fig:Noise_results.jpg}c, g, and k, the inversion results with noise at $std=0.5$ for $V_P$, $V_S$, and ${\rho}$ are plotted; in Figure \ref{fig:Noise_results.jpg}d, h, and l are likewise for $std=1.0$.  We conclude that a moderate amount of random error in the data used for the RNN training leads to acceptable results, though some blurring is introduced and detail in the structure is lost.  The $V_S$ recovery appears to be much more sensitive to noise than are those of $V_P$ or ${\rho}$. 

\section{Conclusions}

Elastic multi-parameter full waveform inversion can be formulated as a strongly constrained, theory-based deep-learning network.  Specifically, a recurrent neural network, set up with rules enforcing elastic wave propagation, with the wavefield projected onto a measurement surface acting as the labeled data to be compared with observed seismic data, recapitulates elastic FWI but with both (1) the opportunity for data-driven learning to be incorporated, and (2) a design supported by powerful cloud computing architectures. Each cell of the recurrent neural network is designed according to the isotropic-elastic wave equation.       

The partial derivatives of the data residual with respect to the trainable parameters which act to represent the elastic media are calculated by using the intelligent automatic differential method. With the automatic differential method, gradients can be automatically calculated via the chain rule, guided by  backpropagation along the paths within the computational graph. The automatic differential method produces a high level of computational efficiency and scalability for the calculation of gradients for different parameters in elastic media.  The formulation is suitable for exploring numerical features of different misfits and different parameterizations, with an aim of improving the resolution of the recovered elastic models, and mitigate cross-talk. 

We compared RNN waveform inversions based on $l_{2}$, $l_{1}$ with high order total variations. We used this RNN synthetic environment to compare density-velocity (P-wave velocity, S-wave velocity, and density), modulus-density (Lam\'{e} parameters and density) and S-D ($C_{11}$, $C_{44}$ and density) parameterizations, and their exposure to cross-talk for the varying misfit functions. Our results suggest generally that this approach to full waveform inversion is consistent and stable. S-D $l1$ norm with high order TV regulation can better resolve the elastic synthetic models.   $l1$ norm with high order TV regulation can mitigate the cross-talk between parameters with gradient-based method.



\bibliographystyle{seg}  
\bibliography{Tzhang.bib}

\end{document}